\documentclass[aps,pre,twocolumn,showpacs,groupedaddress,lengthcheck]{revtex4-1}

\pdfoutput=1
\usepackage[utf8]{inputenc} 			
\usepackage[T1]{fontenc}				
\usepackage{lmodern}					
\usepackage{amsmath,amssymb}
\usepackage{graphicx}
\usepackage{float}
\usepackage[usenames,dvipsnames]{xcolor}
\usepackage{tikz}
\usepackage{hyperref}

\newcommand{\eps}{\varepsilon}

\newcommand\cP{{\cal P}}
\newcommand\cN{{\cal N}}

\newcommand\grad{\nabla}

\renewcommand\vec[1]{{\bf #1}}

\begin{document}

\title{Flocking with discrete symmetry: the 2d Active Ising Model}

\author{A. P. Solon, J. Tailleur}
\affiliation{Universit\' e Paris Diderot, Sorbonne Paris Cit\' e, MSC, UMR 7057 CNRS, F75205 Paris, France}

\date{\today}
 
\begin{abstract}
  We study in detail the active Ising model, a stochastic lattice gas
  where collective motion emerges from the spontaneous breaking of a
  discrete symmetry. On a 2d lattice, active particles undergo a
  diffusion biased in one of two possible directions (left and right)
  and align ferromagnetically their direction of motion, hence
  yielding a minimal flocking model with discrete rotational
  symmetry. We show that the transition to collective motion amounts
  in this model to a \textit{bona fide} liquid-gas phase transition in
  the canonical ensemble. The phase diagram in the density/velocity
  parameter plane has a critical point at zero velocity which belongs
  to the Ising universality class. In the density/temperature
  `canonical' ensemble, the usual critical point of the equilibrium
  liquid-gas transition is sent to infinite density because the
  different symmetries between liquid and gas phases preclude a
  supercritical region. We build a continuum theory which reproduces
  qualitatively the behavior of the microscopic model. In particular
  we predict analytically the shapes of the phase diagrams in the
  vicinity of the critical points, the binodal and spinodal densities
  at coexistence, and the speeds and shapes of the phase-separated
  profiles.
\end{abstract}

\maketitle

\tableofcontents
\section{Introduction}
Active matter systems, defined as large assemblies of interacting
particles consuming energy to self-propel, exhibit a variety of
elaborate collective behaviors. Among them, collective motion---a term
referring to the coherent displacement of large groups of individuals
over length scales much larger than their individual size---has played a
leading role in active matter studies.  It can be observed in a wide
range of biological systems such as bird flocks~\cite{Ballerini2008},
fish schools~\cite{fish,fishHugues}, bacterial
swarms~\cite{Steager2008,PeruaniMB}, actin~\cite{Schaller2010} or
microtubule~\cite{SuminoNature2010} motility assays, but also in inert
matter that is artificially self-propelled, for example in assemblies
of vibrated polar disks~\cite{Deseigne2010}, rolling
colloids~\cite{BartoloNature} or self-propelled liquid
droplets~\cite{Thutupalli}.

On the theoretical side, the transition to collective
motion---hereafter referred to as the ``flocking'' transition---has
attracted the attention of the physics community because simple models
have proved useful to describe its generic properties, highlighting
the possibility of universal behaviours. The model introduced by
Vicsek and collaborators two decades ago~\cite{Vicsek1995} is
prototypical of this line of research, containing only two
ingredients: self-propulsion at a constant speed and aligning
interactions.  It has often been described as a dynamical XY
model~\cite{TT} since the alignment of the particle directions of
motion resemble the ferromagnetic alignment of XY spins.

The phenomenology of the Vicsek model is now well
established~\cite{Vicsek1995,Gregoire2004,SolonVMPRL2015}. When
decreasing the noise on the aligning interaction, or increasing the
density, a transition takes place from a disordered gas into an
ordered state of collective motion. Between these two homogeneous
phases lays a region of parameter space where particles gather in
dense ordered bands travelling in a dilute disordered
background. These bands, which are a robust feature of flocking
models~\cite{Gregoire2004,Mishra2010,Weber2013,PeruaniGliders,Farrell2012,BDG,Ihle},
are a signature of the first-order nature of the transition, together
with intermittency, metastability and
hysteresis~\cite{Gregoire2004,SolonVMPRL2015}. Unfortunately, they are
seen only in large systems and strong finite size effects render the
numerical study of the Vicsek model (VM) very costly in computing
power.

To overcome these numerical difficulties and gain more insight into
the flocking transition, a number of analytical approaches have been
followed. Hydrodynamic equations for Vicsek-like models have been
either derived by coarse-graining~\cite{BDG,Ihle} or proposed
phenomenologically~\cite{TT,Ramaswamy2005,Mishra2010}. These equations
predict phase diagrams in qualitative agreement with the microscopic
models, including the existence of inhomogeneous bands~\cite{BDG,Ihle,
  Mishra2010,SolonVMPRL2015}. Their analytical study is however so
complicated that little can be done beyond working with their
linearized version. Nevertheless, some progress was made to account
for the long range order and giant density fluctuations observed in
the ordered phase of the Vicsek model~\cite{TT}. Interestingly, it was
also recently shown that all hydrodynamic equations derived for polar
flocking models~\cite{BDG,Mishra2010,Ihle,Solon2013} admit the same
family of 1d propagative solutions~\cite{Caussin}. A complete
analytical study of the Vicsek model, from micro to macro, however
remains elusive.

An alternative strategy to gain insight into the flocking transition
relied on the introduction of an Active Ising Model
(AIM)~\cite{Solon2013} which circumvents both the numerical and
analytical pitfalls of the Vicsek model. Using non-equilibrium
versions of ferromagnetic models has indeed often proven a useful
strategy~\cite{DDS,PeruaniGliders,KimPRE2015,Eaves}. The AIM, which we study
in detail in this paper, contains the two key ingredients for
flocking: self-propulsion and aligning interactions. The continuous
rotational symmetry of the Vicsek model is however replaced by a
discrete symmetry; In the AIM, particles diffuse in the 2d plane but
are self-propelled in only one of two possible directions (left or
right). It is thus akin to a dynamical Ising model where particles
have a discrete rotational symmetry. The AIM is found to have a
simpler, more tractable, behavior than the Vicsek-like models with
continuous symmetry while still retaining a large part of their
physics. Using a lattice-based model also simplifies both numerical
and analytical studies.

After introducing the model in section~\ref{sec:def}, we present a
numerical study of the 2d AIM in section~\ref{sec:phase_diagrams}. Our
main conclusion is that the transition in the AIM amounts to a
liquid-gas transition in the canonical ensemble. At fixed orientational
noise, the system can be in two ``pure'' states: a disordered gas or
an ordered liquid, the latter leading to a collective migration of all
particles to the left or to the right. When constraining the system's
density to lie between two 'spinodal lines', no homogeneous phase can
be observed and the system phase-separates, with an ordered travelling
liquid band coexisting with a disordered gas background. A key
difference with the usual equilibrium liquid-gas transition is that
liquid and gas have different symmetries; A supercritical region is
thus prohibited since one has to break a symmetry to take the system
from a gas to a liquid state, which explains the atypical shape of the
phase diagram.

In section~\ref{sec:hydro}, we complement our numerical approach by
deriving a set of hydrodynamic equations for the dynamics of the local
density and magnetisation fields. Interestingly, a simple mean-field
theory wrongly predicts a continuous transition, failing to account
for the phase-separated profiles. A refined mean-field model, taking
into account the fluctuations of the density and magnetisation fields,
reproduces qualitatively the phenomenology of the AIM. In
section~\ref{sec:fronts}, we use the hydrodynamic equations to compute
at large densities the shape of the phase-separated profiles, the
coexisting densities, the velocity of the liquid domain and account
for the finite-size scalings observed in the microscopic
model. Finally, we argue in section~\ref{sec:robustness} in favor of
the robustness of our results by considering an off-lattice version of
the model and different boundary conditions.
\begin{figure}
\centering
\includegraphics{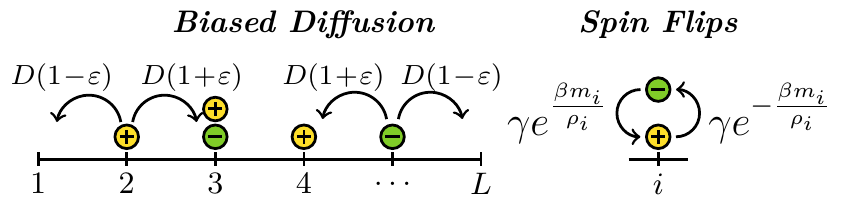}
  \caption{Sketch of the two possible actions and their rates of
    occurrence. The ferromagnetic interaction between particles is
    purely on-site and particles diffuse freely. Beyond the biased
    diffusion shown here, particles also hop symmetrically up or down,
    with equal rates $D$ in both directions.}
  \label{fig:rates}
\end{figure}

\section{Definition of the model}
\label{sec:def}
We consider $N$ particles moving on a 2D lattice of $L_x\times L_y$
sites with periodic boundary conditions. Each particle carries a spin
$\pm 1$ and there are no excluded volume interactions between the
particles: there can thus be an arbitrary number $n_i^{\pm}$ of
particles with spins $\pm 1$ on each site $i\equiv (i_1,i_2)$. The
local densities and magnetizations are then defined as
$\rho_i=n^+_i+n^-_i$ and $m_i=n^+_i-n_i^-$.  We consider a
continuous-time Markov process in which particles can both flip their
spins and hop to neighboring lattice sites at rates that depend on
their spins (see Fig.~\ref{fig:rates}). The hopping and flipping
rates, detailed in the next subsections, are such that our model is
endowed with self-propulsion and inter-particle alignment, hence
consituting a flocking model with discrete symmetry.

\subsection{Alignment: Fully connected Ising models}
A particle with spin $s$ on site $i$ flips its spin at rate
\begin{equation}
  \label{eq:rates}
W(s\to-s)=\gamma \exp\left(-s\beta\frac{m_i}{\rho_i}\right),
\end{equation}
where $\beta=1/T$ plays the role of an inverse temperature. These
rates satisfy detailed balance with respect to an equilibrium
distribution $P\propto \exp[-\beta H]$ where $H$ is the sum over the
$L_x L_y$ lattice sites of the Hamiltonians of fully connected Ising
models:
\begin{equation}
  \label{eq:hamiltonian}
  H = - \sum_{\text{sites }i} \frac{1}{2 \rho_i} \sum_{j=1}^{\rho_i} \sum_{k \neq j} S^i_jS^i_k=-\sum_{\text{sites }i}\left[ \frac{m_i^2}{2\rho_i}-\frac{1}{2}\right]
\end{equation}
The first sum runs over the lattice site index $i=(i_1,i_2)$, the next
two over the particles $j,k$ present on site $i$, and $S^i_j=\pm 1$ is
the value of spin $j$.  (The factor $1/2$ simply avoids double
counting.) The rate $\gamma$ can always be absorbed in a change of
time unit so that we take $\gamma=1$, silently omitting it from now
on.

This interaction is purely local: particles only align with other
particles on the same site and, without particle hopping, the model
amounts to $L^2$ independent fully connected Ising models. The factor
$1/\rho_i$ in $W$ makes the Hamiltonian $H$ extensive with $N$ and
keeps the interaction rates bounded: the rate $W(s\to -s)$ at which a
particle of spin $s$ flips its spin varies between $\exp(-\beta)$ if
all the other particles on the same site have spins $s$ to
$\exp[\beta(1-2/\rho_i)]$ if they all have spins $-s$.

\begin{figure}
  \centering
  \includegraphics{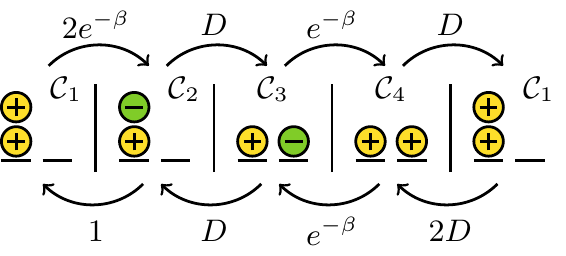}
  \caption{A loop of four configurations involving 2 particles on 2
    sites breaking Kolmogorov's criterion\cite{Kolmo} showing that the system does
    not satisfy detailed balance even when $\eps=0$. The numbers
    associated to the arrows are the transition rates for
    $\eps=0$. The product of the transition rates along
    $\mathcal{C}_1\to\mathcal{C}_2\to\mathcal{C}_3\to\mathcal{C}_4\to\mathcal{C}_1$
    (left to right) is $2 D^2 e^{-2\beta}$, whereas the reverse order
    (right to left) yields $2 D^2 e^{-\beta}$.}
  \label{fig:Kloop}
\end{figure}

\begin{figure}
  \includegraphics[width=\columnwidth]{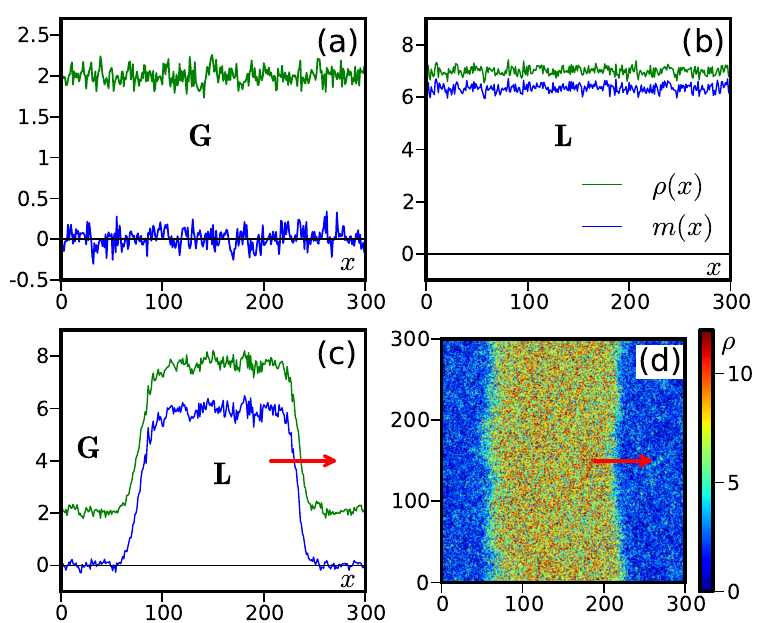}
  \caption{(a-c) Examples of density profiles (green upper line) and magnetization
    profiles (blue lower line) averaged along vertical direction for the
    three phases. (a) Disordered gas, $\beta=1.4$, $\rho_0=2$. (b)
    Polar liquid, $\beta=2$, $\rho_0=7$. (c) Liquid-gas coexistence,
    $\beta=1.6$, $\rho_0=5$. (d) 2d snapshot corresponding to
    (c). (for all figures $D=1$, $\eps=0.9$)}
  \label{fig:phases}
\end{figure}
\subsection{Self-propulsion: Biased diffusion}
Particles also undergo free diffusion on the lattice, with a left/right
bias depending on the sign of their spins: a particle with spin $s$
hops with rate $D(1+s\eps)$ to its right, $D(1-s\eps)$ to its left,
and $D$ in both the up and down directions. There is thus a mean drift,
which plays the role of self-propulsion, with particles of spins $\pm
1$ moving along the horizontal axis with an average velocity $\pm
2D\eps$.

The model is designed to have the self-propulsion entering in a
minimal and tunable way through the parameter $\eps$. Importantly, the
limit of vanishing self-propulsion $\eps\to 0$ is well-defined because
the spins still diffuse on the lattice. This dynamics should thus
allow us to interpole continuously between `totally self-propelled'
($\eps=1$), self-propelled ($\eps\in]0,1[$), weakly self-propelled
($\eps\sim 1/L$) and purely diffusive $(\eps=0)$ particles.

This differs from the Vicsek model where the zero-velocity limit
corresponds to immobile particles undergoing an equilibrium dynamics
resembling that of the XY model, with a quenched disorder on the bonds
(only particles closer than a fixed distance interact).

Let us note, however, that even when $\eps=0$ the model is not at
equilibrium {\it i.e.}  it does not satisfy detailed balance with
respect to any distribution. This is easily shown using Kolmogorov's
criterion~\cite{Kolmo}. In Fig.~\ref{fig:Kloop}, we exhibit a loop of four
configurations such that the products of the transition rates for
visiting the loop in one order,
$\mathcal{C}_1\to\mathcal{C}_2\to\mathcal{C}_3\to\mathcal{C}_4\to\mathcal{C}_1$,
and the reverse order are different, whence a violation of detailed
balance. To make the $\eps=0$ limit an equilibrium dynamics, one
strategy could be to choose hopping rates satisfying detailed balance
with respect to the Hamiltonian H defined in~\eqref{eq:hamiltonian},
replacing $D$ by $D \exp(-\beta \Delta H/2)$). The steady-state
distribution would however be factorized and not very interesting. An
alternative would be to further add to~\eqref{eq:hamiltonian} nearest
neighbours interactions but we have not followed this cumbersome path
here. Actually, as we show in section~\ref{sec:phase_diagrams_epsrho},
this microscopic irreversibility when $\eps=0$ is irrelevant at large
scales and we recover in this limit a phase transition belonging to
the Ising universality class.

\subsection{Simulations}
To simulate the dynamics of the model, we used a
random-sequential-update algorithm. We discretized the time in small
time-steps $\Delta t$. A particle is then chosen at random; it flips
its spin $s$ with probability $W(s\to -s) \Delta t$, hops upwards or
downwards with probabilities $D\Delta t$, to its right or to its left
neighboring sites with probabilities $D(1\pm s \eps)\Delta
t$. Finally, it does nothing with probability $1-[4D+W(s\to
-s)]\Delta t$. Time is then incremented by $\Delta t/N$ and we iterate
up to some final time. In practice we used $\Delta t=[4
D+\exp(\beta)]^{-1}$ to minimize the probability that nothing happens
while keeping all probabilities smaller than one. 

Note that this algorithm does not allow a particle to be updated twice
(on average) during $\Delta t$ and is thus an approximation of our
continuous-time Markov process. We also used continuous-time
simulations, associating clocks to each particle or each site and
pulling updating times from the corresponding exponential laws. In
practice we did not find any difference in the simulation results but
the continuous time simulations were often slower so that we mostly
used the random sequential update algorithm.

In most of this article we use simulation boxes with $L_x\times L_y$
lattice sites and periodic boundary conditions. In
section~\ref{sec:closed} we discuss what happens for closed boundary conditions.

\begin{figure}
  \includegraphics{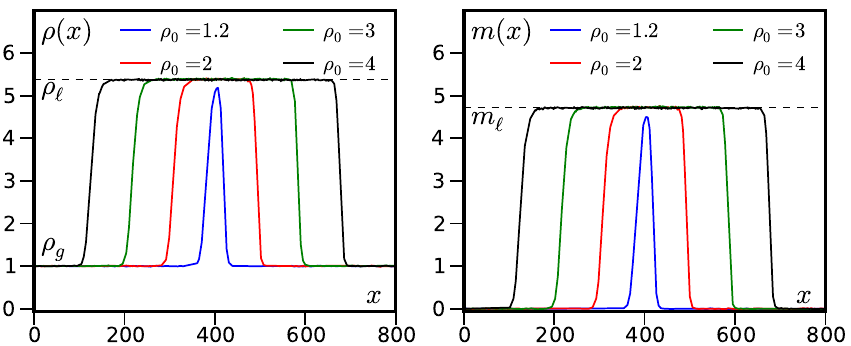}
  \caption{Phase-separated density (left) and magnetization (right)
    profiles as the density increases. Parameters: $\beta=2$, $D=1$,
    $\eps=0.9$, system of size 800x100. The profiles have been
    averaged over time and along the $y$ axis.}
  \label{fig:profiles_micro}
\end{figure}

\begin{figure}
  \includegraphics[width=1\columnwidth]{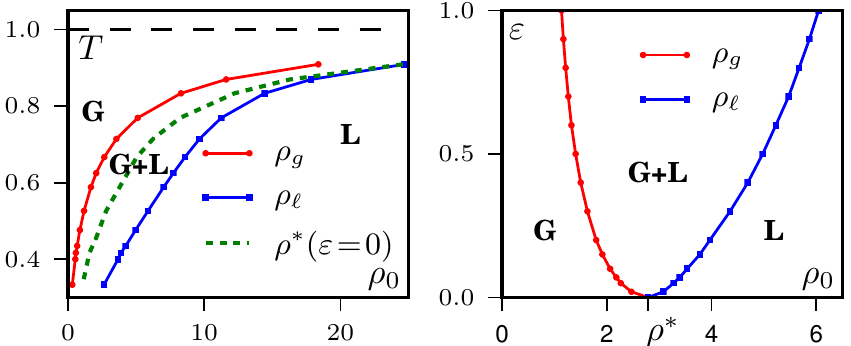}
  \caption{Phase diagrams of the AIM. The red and blue lines
    delimit the region of existence of phase-separated profiles.
    {\bf Left}: Parameter spaces ($T=1/\beta$, $\rho_0$) for
    $D=1$. Red and blue coexistence lines correspond to $\eps=0.9$
    while the green dashed line indicates the critical points at
    $\eps=0$. {\bf Right}: Parameter space ($\eps$, $\rho_0$) for
    $D=1$, $\beta=1.9$. At $\eps=0$ we recover critical point in the
    Ising universality class.}
  \label{fig:phase_diagrams}
\end{figure}
\begin{figure}
  \includegraphics{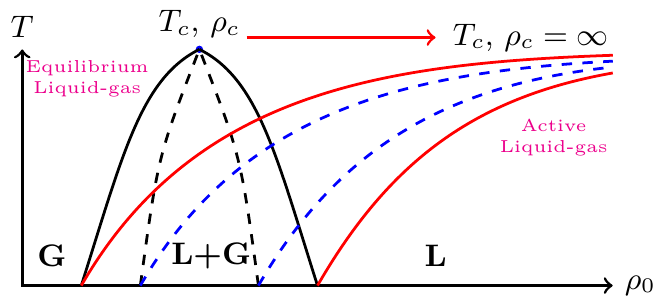}
  \caption{Schematic picture of the differences between the phase
    diagrams of the passive and active liquid gas transition. In the
    active case, because the liquid and the gas have different
    symmetries, the critical point is sent to $\rho=\infty$, thus
    suppressing the supercritical region.}
  \label{fig:passive_active_liquidgas}
\end{figure}

\section{A Liquid-Gas Phase Transition}
\label{sec:phase_diagrams}
We explored the phase diagram using three control parameters: the
temperature $T= \beta^{-1}$, the average density $\rho_0=N/(L_xL_y)$,
and the self-propulsion `speed' $\eps$. Doing so, we observed three
different phases shown in Fig.~\ref{fig:phases}. For $\eps\neq 0$, at
high temperatures and low densities, the particles fail to organize
and we observe a homogeneous gas of particles with local magnetization
$\langle m_i \rangle\approx 0$. On the contrary, for large densities
and small temperatures, the particles move collectively either to the
right or to the left, forming a polar liquid state with $\langle
m_i\rangle = m_0 \neq 0$. For intermediate densities, when $\rho_0 \in
[\rho_g(T,\epsilon),\,\rho_\ell(T,\epsilon)]$, the system phase
separates into a band of polar liquid traveling to the left or to the
right through a disordered gaseous background.

The lines $\rho_g(T,\epsilon)$ and $\rho_\ell(T,\epsilon)$ both
delimit the domain of existence of the phase-separated profiles and
play the role of coexistence lines: As shown in
Fig.~\ref{fig:profiles_micro}, for all phase-separated profiles at
fixed $T,\,\epsilon$, the densities in the gas and liquid part of the
profiles are $\rho_g$ and $\rho_\ell$, respectively. Correspondingly,
the magnetization are $0$ and $m_\ell(T,\epsilon)\neq 0$. Thus, varying
the density $\rho_0$ at constant temperature and propulsion speed
solely changes the width of the liquid band. Consequently, in the
phase coexistence region, the lever rule can be used to determine the
liquid fraction $\Phi$ in the same way as for an equilibrium
liquid-gas phase transition in the canonical ensemble:
\begin{equation}\label{eqn:phivsrho}
  \Phi = \frac{\rho_0-\rho_g}{\rho_\ell-\rho_g}
\end{equation}
As we shall see below, this analogy goes beyond the sole shape of the
phase-separated profiles and the phase-transition to collective motion
of the active Ising model is best described as a liquid-gas phase
transition rather than an order-disorder one.

\subsection{Temperature-density `canonical' ensemble}
The phase diagram in the ($T$, $\rho_0$) parameter plane, computed for
$\eps=0.9$, is shown in the left panel of
Fig.~\ref{fig:phase_diagrams}. While the general structure of the
phase diagram, with a gas phase, a liquid phase, and a coexistence
region, is reminiscent of an equilibrium liquid-gas phase diagram, the
shapes of the transition lines are unusual. This difference can be
understood using a symmetry argument. Since the disordered gas and the
polar liquid have different symmetries, the system cannot continuously
transform from one homogeneous phase to the other without crossing a
transition line. There is thus no super-critical region and the
critical point is sent to $T_c=1$ and $\rho_c=\infty$. (See
Fig.~\ref{fig:passive_active_liquidgas} for a schematic picture.)

This symmetry argument should be rather general for flocking
transitions separating a disordered state and a symmetry-broken state
of collective motion. Indeed, in Vicsek-like models, where the role of
the inverse temperature is played by the noise intensity, the phase
diagrams are qualitatively similar to the one shown in
Fig.~\ref{fig:phase_diagrams}. This is true both for the full phase
diagram recently computed in~\cite{SolonVMPRL2015} as well as for earlier
results~\cite{Gregoire2004}, for a slightly different kinetic model
and its hydrodynamic theory~\cite{BDG}, but also for an active nematic
Vicsek-like model~\cite{Sandrine} and a hydrodynamic theory of
self-propelled rods~\cite{Peshkov2012}. 

\subsection{Velocity-density ensemble}
\label{sec:phase_diagrams_epsrho}
Conversely, one can change the strength of the self-propulsion $\eps$
while keeping the temperature fixed. Again, one obtains a phase
diagram with three regions. The difference with the canonical ensemble
is that in this parameter plane, the two coexistence lines merge at
$\eps=0$, where self-propulsion vanishes, yielding a critical point at
a finite density $\rho^*(T)$ (See the right panel of
Fig.~\ref{fig:phase_diagrams}). The curve $\rho^*(T)$ is reported in
the left panel of Fig.~\ref{fig:phase_diagrams} and satisfies
$\rho^*(T)\in [\rho_g(T,\eps),\rho_\ell(T,\eps)]$. In
section~\ref{sec:eps0cp} we show that this critical point belongs to
the Ising universality class.

The shape of this phase diagram is identical to the one computed
in~\cite{Mishra2010} for a phenomenological hydrodynamic description
of self-propelled particles with polar alignment. The comparison with
other microscopic models in the literature is however hard to make
since there seems to be very few studies in the $(\eps,\rho_0)$ plane,
probably because very few models admit a well-defined zero velocity
limit.

\subsection{Nucleation vs spinodal decomposition}
As for an equilibrium liquid-gas transition, the coexistence lines
$\rho_g(T,\eps)$ and $\rho_\ell(T,\eps)$ are complemented by spinodal
lines $\varphi_g(T,\eps)$ and $\varphi_\ell(T,\eps)$ that mark the limit of
linear stability of the homogeneous gas and liquid phases,
respectively. While $\rho_g$ and $\rho_\ell$ are easily measured in
simulations, $\varphi_g$ and $\varphi_\ell$ are much harder to access
numerically at non-zero temperature: When the system is in the
coexistence region but outside the putative spinodal lines, the
homogeneous phases are metastable and finite fluctuations make the
system phase-separate. The closer to the spinodal line, the faster
this nucleation occurs and it is then difficult to pinpoint precisely
the transition from a `fast' nucleation to a spinodal
decomposition. Nevertheless, the differences between the coexistence
and spinodal regions are clearly seen when, starting from a
homogeneous phase, one quenches the system in the coexistence region
but relatively far away from the spinodal lines.

Quenching outside the spinodal region, the homogeneous phases are
metastable. The closer to the binodals, the longer it takes
for a liquid (resp. gas) domain to be nucleated in the gas
(resp. liquid) background. The convergence to the phase-separated
steady-state then results from the coarsening of this domain.
\begin{figure}
  \includegraphics[width=\columnwidth]{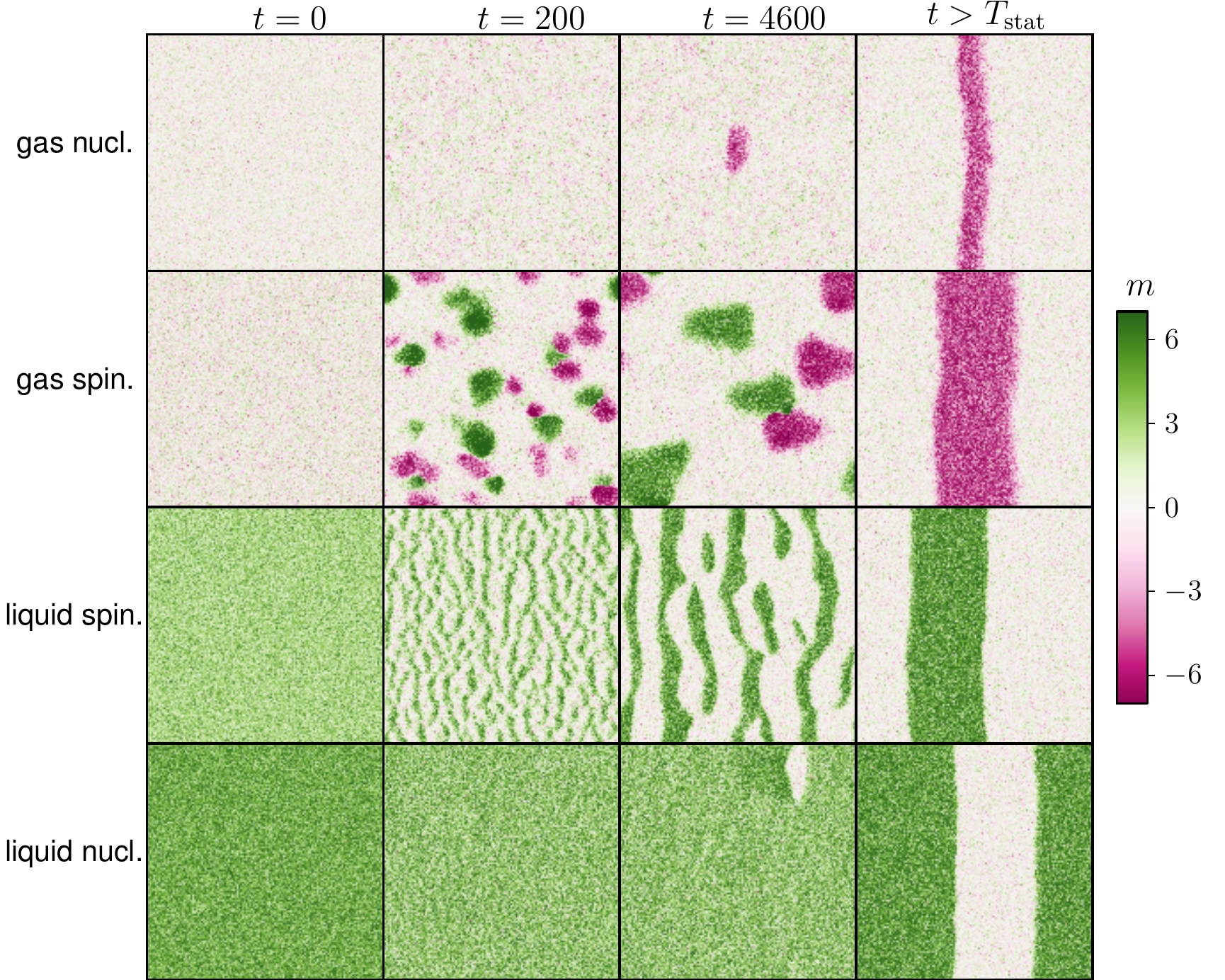}
  \caption{Successive snapshots following quenches from homogeneous
    gas and liquid phases inside and outside the spinodal
    region. Parameters: $D=1,\eps=0.9,\beta=1.8$, system sizes 400x400
    and 1000x1000 for the quenches from the gas and liquid
    phases. From top to bottom, $\rho_0=1.84,\,3,\,3,\,4.7$. See
    Supplementary Movies in~\cite{SI}.}
\label{fig:quench_snap}
\end{figure}

\begin{figure}
  \includegraphics[width=\columnwidth]{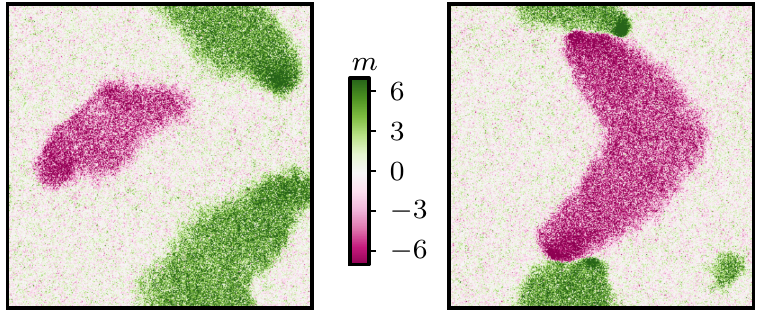}
  \caption{Snashots in the late stage of coarsening taken from the
    same simulation as the first row of fig.~\ref{fig:quench_snap} at
    time $t=283000$ (left) and $t=310000$ (right). Parameters:
    $D=1,\eps=0.9,\beta=1.8,\rho_0=3$, system sizes 400x400.}
\label{fig:quench_snap_late}
\end{figure}

Quenching inside the spinodal region, the different symmetries between
gas and liquid result in different spinodal decomposition dynamics
when starting from ordered and disordered phases. Starting from a
disordered gas, the linear instability almost immediately results in
the formation of an extensive number of small clusters of negative and
positive spins. The coarsening then stems from the merging of these
clusters, until a single, macroscopic domain remains. The late stage
of the coarsening is thus dominated by the long-lived competition
between a small number of right- and left-moving macroscopic
clusters. Their shapes (see Fig.~\ref{fig:quench_snap_late}) are
reminiscent of the counter-propagating arrays of bands reported
in~\cite{FreyPRX}, where it was suggested, using deterministic
simulations of the Boltzmann equation derived for kinetic flocking
models, that such profiles could constitute a new phase of flocking
models. In our simulations, we always observed a coarsening process
leading to a single band, which seems to indicate that the apparent
stability of these solutions in~\cite{FreyPRX} could be due to
the lack of fluctuation terms. It would nevertheless be interesting to
make a more detailed study of the coarsening dynamics to see if these
alternating bands could indeed form a stable phase (for instance at
low temperatures, where the coarsening seems to become slower and
slower).

Starting from the ordered phase, the linear instability results in
many liquid domains which all move in the same direction. The
coarsening then results from the collision of liquid bands that move
in the same direction, but with slightly different speeds.  See
Fig.~\ref{fig:quench_snap} and SI movies~\cite{SI} for examples of
these four possible dynamics.

\subsection{Hysteresis loops}
\label{sec:hysteresis}

\begin{figure}
  \includegraphics{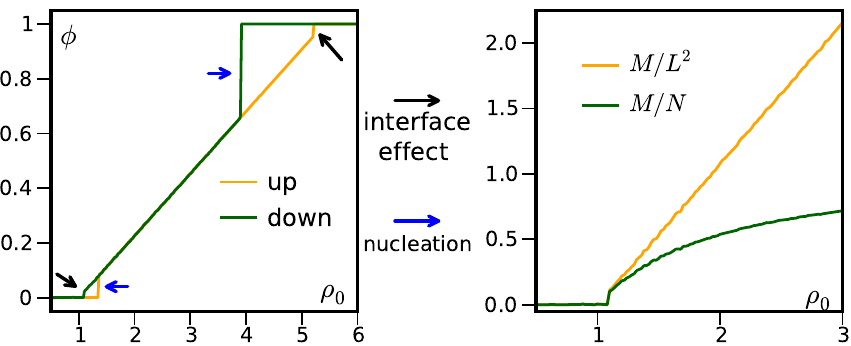}
  \caption{{\bf Left:} Evolution of the liquid fraction $\phi$ upon
    changing continuously $\rho_0$. Large jumps in $\Phi$ correspond
    to the nucleation of bands in meta-stable homogeneous profiles
    while small jumps are finite-size effects due to the finite width
    of the interfaces connecting the gas and liquid regions. The
    density is increased by $\delta \rho_0=0.02$ every $\Delta
    t=2000$. $L_x\times L_y=800\times 100, \,\beta=2, \,\eps=0.9,
    \,D=1$. {\bf Right} Evolution of the magnetisations per particles
    and per sites as the density is varied. The linear scaling typical
    of liquid-gas phase transition is seen using $m_L=M/L^2$.}
  \label{fig:hysteresis_micro}\label{PvsW}
\end{figure}

Another similarity with a liquid-gas transition is the presence of
hysteresis loops obtained by varying slowly the density at constant
$\beta$ and $\eps$ in finite-size systems. Such loops are shown in the
left panel of Fig.~\ref{fig:hysteresis_micro}, where the liquid
fraction $\Phi$ is reported as the density is continuously ramped up
and down. To measure $\Phi$ numerically, a first strategy,
followed in~\cite{Solon2013}, is to compute average density profiles at
fixed $\rho_0$, as shown in the left panel of
Fig.~\ref{fig:profiles_micro} and use an arbitrary density treshold
between $\rho_g$ and $\rho_\ell$ to associate each site to the gas or
liquid regions.

Since the interfaces between gas and liquids are not perfectly
straight, this is slightly artefactual for finite-size systems. Here
we decided to measure $\Phi$ numerically through:
\begin{equation}
\label{eq:phin}
  \Phi_{\rm num.}= \frac 1 {m_\ell L_xL_y} \sum_i m_i
\end{equation}
where $m_\ell$ is the magnetization of the plateau in the liquid part of
the profile. ($m_\ell$ is independent of $\rho_0$ as long as the system
is phase-separated and corresponds to the magnetization of a uniform
liquid phase at the coexistence density $\rho_\ell$.) The results are
very similar to those obtained in~\cite{Solon2013} but first~\eqref{eq:phin} is
faster to measure and second it does not rely on an arbitrary density
treshold.

Starting in the gas phase and increasing the density, the system
remains disordered, with a liquid fraction $\phi=0$, until a band of
liquid is nucleated, at which point $\phi$ jumps to a finite
value. Increasing again $\rho_0$, the liquid region widens until the
two interfaces between gas and liquid almost touch and the liquid phase
{almost} fills the system. At that point, the
system jumps to a homogeneous liquid phase with $\phi=1$.

Upon decreasing the density, a similar scenario occurs: A homogeneous
liquid becomes metastable as the coexistence line is crossed. As the
density keeps decreasing, the system thus remains in a liquid state
with $\Phi=1$ until a nucleation event brings it to a phase-separated
profile. The liquid region then shrinks until its boundaries almost
touch and a second discontinuity of $\Phi$ occurs as the system jumps
into a homogeneous gas phase.

\subsection{Order parameter and finite-size scaling}
The liquid-gas transition picture suggests different finite-size
scaling and order parameter than those associated to magnetic phase
transitions previously used to study flocking models. Most
studies~\cite{Vicsek1997,Martin1999,Gregoire2004} indeed relied on the
mean magnetization per particle
\begin{equation}
   m_N=\frac 1 N \sum_{i} m_i
\end{equation}
rather than the mean magnetization per unit area
\begin{equation}
   m_L = \frac 1 {L_x L_y} \sum_i m_i = \rho_0  m_N 
\end{equation}
For models like the Vicsek model, the former is nothing but the
polarisation $ \vec m_N = \vec P$ while the latter is related to the
total momentum $ \vec m_L = \rho_0 \vec P /v$. In the phase-separated
region, both can be related to the liquid fraction $\Phi$
through~Eq.~\eqref{eqn:phivsrho}
\begin{align}
  m_N&=\frac 1 N \Phi L_x L_y m_\ell= m_\ell \frac{1-\rho_g/\rho_0}{\rho_\ell-\rho_g}\\
  m_L&=\Phi m_\ell =  m_\ell  \frac{\rho_0-\rho_g}{\rho_\ell-\rho_g}
\end{align}
The simple linear scaling of $m_L$ with $\rho_0-\rho_g$ is replaced by
a non-linear dependence of $m_N$ with $\rho_0$, as shown in
Fig.~\ref{PvsW}, right panel. An apparently inoccent change of the
normalization used to make the magnetisation $M=\sum_i m_i$ intensive
can thus turn the simple affine scaling of $m_L$ with $\rho_0$,
typical of a liquid-gas transition, into the non-linear dependence of
$m_N$ that could make one mistake the transition for a critical one.

\begin{figure}
  \includegraphics{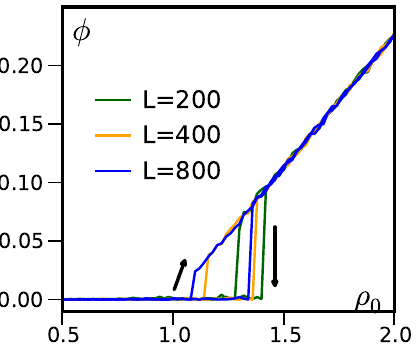} \includegraphics{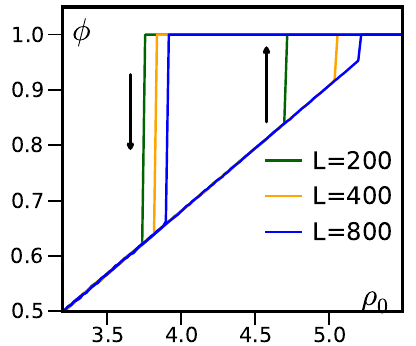}
\caption{Hysteresis loops for system sizes 200x100, 400x100 and
  800x100. For each system, the density is increased by $\delta
  \rho_0=0.02$ every $\Delta t=2000$. $T=0.5$, $\eps=0.9$.}
\label{fig:FSS}
\end{figure}

Let us now go back to the hysteresis loops and discuss their finite
size scaling. As shown in figure~\ref{fig:FSS}, the discontinuities of
the liquid fraction get closer and closer to the binodals $\rho_g$ and
$\rho_\ell$ as the system size increases, leading to vanishingly small
hysteresis loops in the thermodynamic limit.

\begin{figure}
  \includegraphics[width=1\columnwidth]{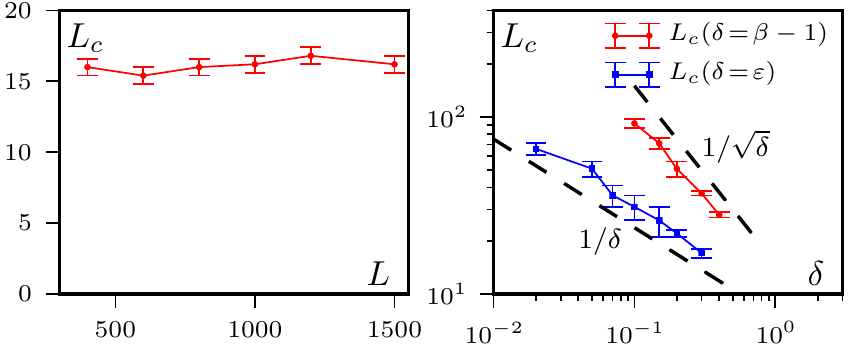}
  \caption{{\bf Left:} Divergence of the critical nucleus $L_c$ when
    approaching the critical points $\beta\to 1$ and $\eps\to 0$. To
    measure $L_c$, we started in the phase-separated state and
    decreased continuously the density (the errorbars correspond to
    the density step used) to record the density $\rho_m$ at which the
    liquid band disappears. $L_c$ is then defined by
    $L\rho_m=L\rho_g+L_c(\rho_\ell-\rho_g)$, as the length of a band
    at density $\rho_\ell$ that can be made with the excess density
    $\rho_m-\rho_g$. {\bf Right:} Variation of the critical nucleus
    iwht $L$ showing that, within numerical errors, it does not depend
    on system size. Parameters: $D=1$, $\eps=0.9$, $\beta=1.9$}
  \label{fig:critical_nucleus}
\end{figure}

Consider first the transition from gas to phase-separated
profiles. The liquid fraction exhibits two different discontinuities
when the density is decreased or increased, due to two different
effects. As the density is decreased, phase-separated profiles cannot
be maintained arbitrarily close to $\rho_g$. There is indeed a
critical nucleus, which roughly amounts to two connected domain walls,
as can be seen in Fig.~\ref{fig:profiles_micro} for $\rho_0=1.2$. As
shown on Fig.~\ref{fig:critical_nucleus}~(left), this critical nucleus
$L_c$ is independant of the system size. If the excess mass $L_x L_y
(\rho_0-\rho_g)$ is smaller than a critical value $\varphi_cL_y$, this
critical nucleus cannot be accomodated in the system, which thus falls
into the gas phase. As the system size increases, the minimal
density to observe phase-separated profiles
$\rho_0=\rho_g+\varphi_c/(L_x)$ thus converges to $\rho_g$ as $L_x$
increases and phase-separated profiles are seen closer and closer to
the binodal.  The second discontinuity, met upon increasing the
density, corresponds to the nucleation of a liquid band of width $L_b$
in a gaseous background. Since $L_b$ can be anything between $L_c$ and
$L_x$, increasing the system size at fixed density should decrease the
mean time until nucleation of such bands, thanks to an entropic
contribution due to the number of places where the bands can be
nucleated. As shown in~\ref{fig:FSS}, this is indeed the case and the
transition to phase-separated profiles thus also happens closer and
closer to the binodal $\rho_g$.

The same line of reasoning can be used to understand the scaling of
the second hysteresis window, close to $\rho_\ell$. Thus, in the
thermodynamic limit, all discontinuities disappear and the liquid
fraction varies continuously from $\phi=0$ at $\rho_0=\rho_g$ to
$\phi=1$ at $\rho_0=\rho_\ell$, as for an equilibrium liquid-gas
transition in the canonical ensemble. Note that the width of the
critical nucleus diverges as one gets closer and closer to the
critical points ($\eps=0$ or $\beta=1$), as shown in the right panel
of Fig.~\ref{fig:critical_nucleus}. This could explain why some
studies of the Vicsek model in the small velocity region claim to find
a critical transition~\cite{AlbanoVM}: as one gets closer and closer to the zero speed
limit, the system-size above which one can correctly observe the
discontinuous nature of the transition diverges.

\begin{figure*}
  \includegraphics[width=.85\textwidth]{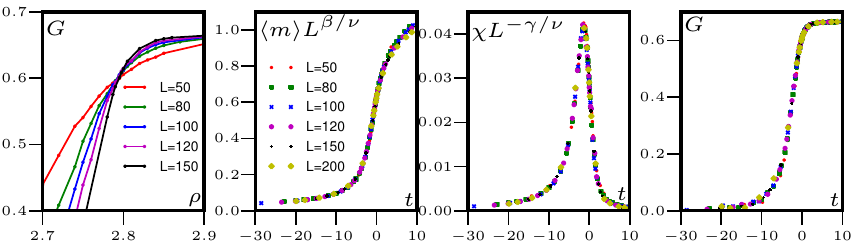}
  \caption{Left: Binder cumulant $G(\rho)$ from which we find the
    critical density $\rho^*=2.798 \pm 0.002$. Other three figures:
    data collapse on the universal scaling functions $F_m$, $F_\chi$
    and $F_G$ (defined in the text) when the data is rescaled with the
    2d Ising exponent values $\beta=1/8$, $\gamma=7/4$ and
    $\nu=1$. $t=L^{1/\nu}(\rho-\rho^*)/\rho^*$. Parameters: $D=1$,
    $\eps=0.9$, $\beta=1.9$.}
  \label{fig:binder_exponents}
\end{figure*}

\subsection{The $\epsilon=0$ critical point}
\label{sec:eps0cp}
While the $\beta=1$, $\rho_c=\infty$, critical point is out of reach
numerically, the study of the $\eps=0$ critical point is
accessible. At $\eps=0$, there is no self-propulsion and the phase
transition is of a completely different nature from the liquid-gas
transition described above. As we show below, despite the dynamics
being non-equilibrium, it turns out to be a standard critical phase
transition belonging to the Ising universality class.

We studied this critical point using a finite-size scaling standard
for magnetic systems at criticality~\cite{Binder1997}. We thus
consider the magnetization $m_N\in [0,1]$. In equilibrium, around
ferromagnetic critical points, the order parameter, susceptibility and
Binder cumulant $G=1-\frac{\langle m^4 \rangle}{\langle m^2
  \rangle^2}$ are known \cite{Binder1997} to obey the finite-size
scaling relations

\begin{align}
  \label{eq:FSS_critical}
  &\langle m \rangle=L^{-\beta/\nu}F_m(tL^{1/\nu}) \\
  &\chi = L^2(\langle m^2 \rangle-\langle m \rangle^2)=L^{\gamma/\nu}F_\chi(tL^{1/\nu}) \\
  & G=F_G(tL^{1/\nu})
\end{align}
where $t=L^{1/\nu}(\rho-\rho^*)/\rho^*$ is the rescaled distance to
the critical density $\rho^*$. $F_m$, $F_\chi$ and $F_G$ are universal
scaling functions and $\beta$, $\gamma$ and $\nu$ the usual critical
exponents.

We used the fact that $G(t=0)$ is independent of $L$ to find the
critical density, which is thus the density where all the curves
$G(t)$ for different system sizes intersect (Fig.~\ref{fig:binder_exponents}, left). We found that the value at the
crossing point is the same universal value $G(t=0)\simeq0.61$ as in the 2d
Ising model~\cite{Kamieniarz}. A very neat data collapse is further observed
for the critical exponents of the 2d Ising model $\beta=1/8$,
$\gamma=7/4$ and $\nu=1$ (see Fig.~\ref{fig:binder_exponents}). We thus conclude that the critical point at $\eps=0$ is indeed in the Ising
universality class.

Note that a direct evalutation of the critical exponents is much
harder than for the equilibrium Ising model. Here, the dynamics is
fixed so one cannot use alternative dynamics like cluster algorithms
to circumvent the problem of critical slowing down.

\subsection{Number fluctuations}
In most flocking models the homogeneous ordered phase exhibits giant
density
fluctuations~\cite{Gregoire2004,huguesMetric,Mishra2010,Dey2012,Sandrine}. These
are quantified by measuring number fluctuations, {\it i.e} by counting
the number of particles $n(\ell)$ in boxes of increasing sizes
$\ell<L$ and computing its root mean square $\Delta n(\ell)$. When the
correlation length ${\cal L}$ is finite, a box of size $\ell \gg {\cal
  L}$ can be divided in $(\ell/{\cal L})^2$ independant boxes. The total
number of particles in the large box is then the sum of independent
identically distributed random variables; The central limit theorem
applies and the probability distribution of $n(\ell)$ tends to a
Gaussian. This yields the ``normal'' scaling $\Delta n\sim
n^{1/2}$. On the contrary, one finds in the Vicsek model the anomalous
scaling $\Delta n\sim n^{0.8}$ \cite{Gregoire2004}.

\begin{figure}
  \includegraphics[width=.7\columnwidth]{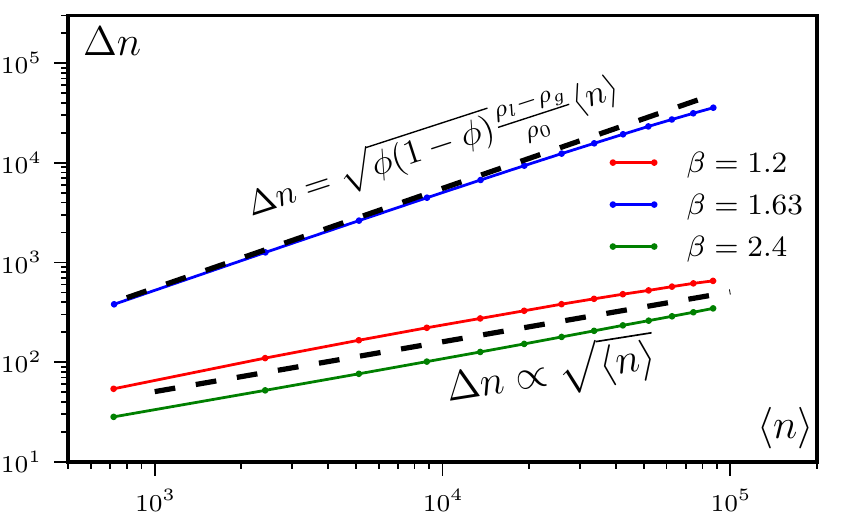}
  \caption{Number fluctuations in the three different phases: gas
    (red), liquid (green) and at coexistence (blue, upper line). $n$
    is the number of particles in boxes of size $\ell$ and $\Delta n$
    its root mean square. $D=1$, $\eps=0.9$, $L=400$, $\rho_0=5$}
  \label{fig:number_fluctuations}
\end{figure}

In the Active Ising model the number fluctuations are found to be
normal in the liquid and gas phases, where $\Delta n\sim n^{1/2}$, and
trivially `giant' in the phase-separated regime where $\Delta n \sim
n$ (see Fig.~\ref{fig:number_fluctuations}).

Note that the scaling $\Delta n\sim n$ is a simple consequence of
phase-separation and one should thus distinguish this scaling from the
`anomalous' scaling of the Vicsek model, which is a signature of
long-range correlations. Let us consider a system at liquid fraction
$\phi$ that is large enough that we can find a range of box sizes
$\ell$ such that: 1)~$\ell\ll L$ so that we can neglect the
contribution of the interfaces (a box is either in the liquid or the
gas phase); 2)~$\ell$ is large enough that $n(\ell)$ takes only two
possible values and we can neglect the fluctuations around these two
values. With these assumptions,
\begin{equation}
  \label{eq:P_n}
  P(n)\simeq \phi \delta(n-\rho_\ell \ell^2)+(1-\phi)\delta(n-\rho_g \ell^2)
\end{equation}
where $\rho_\ell$ and $\rho_g$ are the densities in the gas and liquid
domains. Then one finds
\begin{align}
  \langle n \rangle &=(\phi\rho_\ell+(1-\phi)\rho_g)\ell^2=\rho_0 \ell^2 \\
  \Delta n&=\sqrt{\langle n^2 \rangle-\langle n \rangle^2} =\sqrt{\phi(1-\phi)}\frac{\rho_\ell-\rho_g}{\rho_0} \langle n \rangle
\end{align}
which is a simple hand-waving explanation of the scaling observed in
the coexistence region of the active Ising model, as well as in other
phase-separating systems~\cite{Aranson2008,Fily2012}.

\section{Hydrodynamic description of the Active Ising Model}
\label{sec:hydro}

In this section we derive and analyze a continuous description of
the AIM based on two coupled partial differential
equations accounting for the spatio-temporal evolutions of the density
and magnetization fields. 

We first show in section~\ref{sec:MF} that a standard mean-field
treatment wrongly predicts a continuous transition between the
disordered gas and the ordered liquid. In section~\ref{sec:correct_mf}
we show that local fluctuations, which are neglected in the mean-field
approximations, are necessary to correctly account for the physics of
the system when the density is finite and $\eps\neq 0$. We show in
particular that as soon as the density is finite, fluctuations make
the transition first order. We then use our hydrodynamic description
in section~\ref{sec:fronts} to study the inhomogeneous profiles.

\subsection{Mean-field equations}
\label{sec:MF}
The simplest way to account analytically for a non-equilibrium lattice
gas is probably to derive mean-field equations. These are known to be
quantitatively wrong, but they often capture phase diagrams
correctly~\cite{Blythe2007,Sugden}.

Their derivations follow a standard procedure which can be applied to
the AIM and which, for simplicity, we first present in 1D. Starting from
the master equation, one first derives the time-evolution of the mean
number of $\pm 1$ spins on site $i$
\begin{align}
  \langle \dot n_i^\pm \rangle  &= D (1\pm \eps) \langle n_{i-1}^\pm\rangle  + D(1\mp \eps) \langle n_{i+1}^\pm\rangle  -2 D \langle n_i^\pm \rangle\nonumber \\
  &\pm \langle n_i^{-} \exp(\beta\frac{m_i}{\rho_i})\rangle  \mp \langle n_i^{+} \exp(-\beta\frac{m_i}{\rho_i})\rangle 
\end{align}
which can then be rewritten for the density and magnetisation 
\begin{align}
  \langle \dot \rho_i \rangle  &= D (\langle \rho_{i+1} \rangle +\langle \rho_{i-1} \rangle -2\langle \rho_i \rangle ) - D \eps (\langle m_{i+1} \rangle -\langle m_{i-1} \rangle )\label{eq:rhoav_micro}\\
  \langle \dot m_i \rangle  &= D (  \langle m_{i+1} \rangle +  \langle m_{i-1} \rangle -2   \langle m_i \rangle ) -D \eps (  \langle \rho_{i+1} \rangle -  \langle \rho_{i-1} \rangle )\nonumber\\
  & + 2   \langle \rho_i  \sinh(\beta \frac{  m_i}{ \rho_i})\rangle -2 \langle m_i \cosh(\beta \frac{ m_i}{ \rho_i})\rangle \label{eq:mav_micro}
\end{align}
One can then take a continuum limit using the rescaled variable
$\tilde x=i/L\in [0,1]$, $\tilde D=D/L^2$, $\tilde v=2D\eps/L$ and use
the Taylor expansion $\rho_{i\pm 1}\equiv \rho(x)\pm L^{-1} \partial_x
\rho(x) +L^{-2} \partial_{xx} \rho(x)/2$. We then obtain equations for
the continuum fields $\rho(x)$, $m(x)$, which are assumed to smoothly
interpolate the discrete occupancies $\rho_i$, $m_i$:
\begin{align}
 \partial_t\langle\rho\rangle &= \tilde D \partial_{\tilde x\tilde x} \langle\rho\rangle - \tilde v \partial_{\tilde x} \langle m\rangle \label{eqn:MFrhoav}\\
  \partial_t \langle m\rangle &= \tilde D \partial_{\tilde x\tilde x}\langle m\rangle -
  \tilde v\partial_{\tilde x} \langle\rho\rangle + \left\langle 2 \rho \sinh\frac{\beta m}{\rho} -2m \cosh\frac{\beta m}{\rho}\right\rangle \label{eqn:MFmav}
\end{align}
In higher dimension, the sole difference is that the diffusive terms
become $\tilde D\Delta \langle \rho\rangle$ and $\tilde D \Delta
\langle m \rangle$ whereas the $\tilde v$ terms still involve solely
$\partial_{\tilde x}$ since the hopping is biased only horizontally.

In practice, to compare microscopic simulations and hydrodynamic
theories it is often easier \textit{not} to rescale space and use a
continuous variable $x=L \tilde x \in[0,L]$ (and hence $D=\tilde D
L^2$ and $v=L \tilde v=2D \eps$). Macroscopic and microscopic
transport parameters are then expressed in the same units and
equations~\eqref{eqn:MFrhoav} and~\eqref{eqn:MFmav} are then valid,
without the tilde variables. This is what we use in the following.

Equations~\eqref{eqn:MFrhoav} and~\eqref{eqn:MFmav} are exact; they
couple the first moments $\langle \rho \rangle$ and $\langle m
\rangle$ to higher moments through the averages of the hyperbolic sine
and cosine functions. Following the standard procedure established for
equilibrium ferromagnetic models, we then make two
approximations. First, we take a mean-field approximation by replacing
$\langle f(\rho,m)\rangle$ by $f(\langle \rho\rangle, \langle m
\rangle)$, for any function $f$. (We then drop the
$\langle\dots\rangle$ notation for clarity.) This amounts to
neglecting both the correlations between density and magnetisation and
their fluctuations. Second, we expand the hyperbolic functions in
power series, up to $m^2/\rho^2$. This further restricts our
description to the case where $m \ll \rho$. We then arrive at the
mean-field equations
\begin{align}
  \dot \rho &= D \Delta \rho - v \partial_x m \label{eqn:MFrho}\\
  \dot m &=  D\Delta m -v \partial_x \rho + 2 m (\beta-1) -\alpha \frac{m^3}{\rho^2}\label{eqn:MFmlin}
\end{align}
where $\alpha=\beta^2(1-\beta/3)$. (For $\beta>3$, one should expand to
higher order to obtain a stabilizing term.)

\begin{figure}
\includegraphics{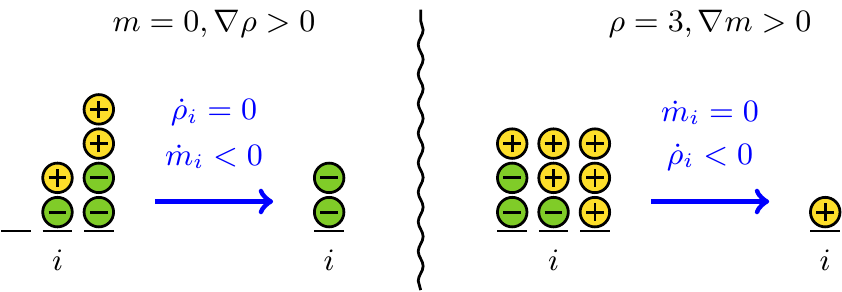}
\caption{Schematic account for the role of density and magnetisation
  gradients in the mean-field equations. {\bf Left}: Initially, $m=0$
  and $\grad \rho>0$ around site $i$. Once plus particles jump to the
  right and minus particles to the left, the density in site $i$ is
  unchanged but $m_i$ has decreased. {\bf Right}: Initially, $\rho$
  is constant and $\grad m<0$ around site $i$. Once particles have
  jumped, the magnetisation of site $i$ is unchanged but the density
  has decreased.}
\label{fig:SPterms}
\end{figure}

Let us consider the various terms appearing in the mean-field
equations. The first terms on the r.h.s of~\eqref{eqn:MFrho}
and~\eqref{eqn:MFmlin} are diffusion terms arising from the stochastic
particle hopping. Let us stress that these terms do \textit{not}
depend on the bias $\eps$ and are thus present even in the totally
asymmetric case $\eps=1$; they do not rely on the possibility for $+1$
and $-1$ particles to hop backwards and forwards, respectively. The
second terms, proportional to $v$, are due to the bias. Their physical
origin is explained in Fig.~\ref{fig:SPterms} where we show how
positive gradients in $m$ or $\rho$ yield negative contributions to
$\dot \rho$ or $\dot m$, respectively. Finally, the last two terms
in~\eqref{eqn:MFmlin} stem from the ferromagnetic interaction and,
apart from the $\rho^2$ dependence of the last term, are typical of
$\phi^4$ Landau mean-field theory. Note that the alignment terms are
the only non-linear ones and thus the only terms for which the
mean-field approximation is actually an approximation.

\begin{figure}
  \includegraphics[width=.9\columnwidth]{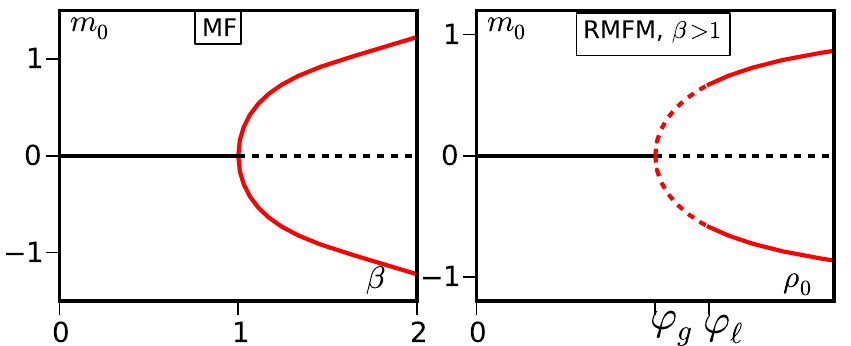}
  \caption{Linear stability of homogeneous profiles in the naive
    (left) and refined (right) mean-field models. Plain (resp. dashed)
    lines denote stable (resp. unstable) solutions. In the RMFM, for
    $\beta<1$, only the homogeneous profile exists and is stable at
    all densities.}
  \label{fig:lin_stab_solutions}
\end{figure}

The mean-field equations always accept the trivial homogeneous
solution
\begin{equation}
  \rho(x)=\rho_0,\qquad m(x)=0\,,
\end{equation}
which is linearly stable for $\beta<1$. As soon as $\beta>1$, two
ordered homogeneous solutions appear,
\begin{equation}
  \rho=\rho_0,\qquad m=\pm\rho_0\sqrt{\frac{2(\beta-1)}\alpha}\,,
\end{equation}
which are linearly stable (see the left panel of
Fig.~\ref{fig:lin_stab_solutions}). Therefore, at the mean-field
level, a linearly stable homogeneous solution exists for all ($\beta$,
$\rho_0$). Furthermore, integrating numerically Eqs.~(\ref{eqn:MFrho})
and (\ref{eqn:MFmlin}) starting from different initial
conditions~\cite{foot1}, the system always relaxes to a homogeneous
solution and inhomogeneous profiles are never observed. Hence, the
mean-field equations predict a continuous transition between
homogeneous disordered and ordered profiles at $\beta=1$, just as for
the Weiss ferromagnet~\cite{Weiss}. The phase diagram is simply split
between a high-temperature disordered homogeneous phase, for $T>1$,
and a low temperature ordered homogeneous phase, for $T<1$.  This
mean-field approach thus completely misses the phenomenology of the
microscopic model; It cannot explain the existence of phase-separated
profiles and yield a phase diagram corresponding to a single
(continuous) transition line at $T=1$, in contradiction to the
crescent shape observed in the microscopic model (see
Fig.~\ref{fig:phase_diagrams}).

\subsection{Going beyond the mean-field approximation}
\label{sec:correct_mf}
Previous coarse-graining approaches of flocking
models~\cite{BDG,Baskaran2008,Farrell2012,Tsimring} often relied on
neglecting correlations by factorizing probability distributions. For
example, in the Boltzmann-Ginzburg-Landau approach of Bertin {\it et
  al.} \cite{BDG} the two-particle probability distribution is
replaced by the product of one-particle distributions. In our
case, to derive the mean-field equations~(\ref{eqn:MFrho}) and
(\ref{eqn:MFmlin}) we made an even cruder approximation.

When computing, for instance, the first non-linear term $\langle m^3/\rho^2
\rangle$, neglecting correlations between $m$ and $\rho$ leads to
\begin{equation}
  \label{eq:approx_corre}
  \langle m^3/\rho^2 \rangle= \langle m^3 \rangle\left\langle \frac{1}{\rho^2} \right\rangle
\end{equation}
We went one step further, completely discarding fluctuations and
replaced $\langle 1/\rho^2 \rangle$, $\langle m^3 \rangle$ by
$1/\langle \rho \rangle^2$, $\langle m \rangle^3$. As we show below,
these fluctuations are crucial to account qualitatively for the
physics of the AIM.

The dynamical equations~\eqref{eqn:MFrhoav} and~\eqref{eqn:MFmav} on
the first moments predict how $\langle\rho(x,t)\rangle$ and $\langle
m(x,t)\rangle$ evolve in time, given an initial distribution
\begin{equation}\label{firstmomIC}
\cP[\rho,m]=\delta(\rho(x)-\rho_0)\delta(m(x)-m_0).
\end{equation}
The mean-field approximation then amounts to compute the averages of
hyperbolic functions in~\eqref{eqn:MFmav} by assuming that, as time goes
on, $\cP$ remains a product of Dirac functions:
\begin{equation}
  \begin{aligned}
  \label{eq:relaxation_MF}
  \cP[\rho,m;x,t \,|\, \rho_0,m_0]&=\delta(\rho(x,t)-\bar \rho(x,t))\\&\quad \times \delta(m(x,t)-\bar m(x,t))
\end{aligned}
\end{equation}
where $\bar\rho(x,t)$ and $\bar m(x,t)$ are solutions of the
mean-field equations~\eqref{eqn:MFrho} and~\eqref{eqn:MFmlin}. In
practice this means that repeatedly simulating the microscopic model
starting from an initial distribution~\eqref{firstmomIC} always yields
the exact same values $\rho(x,t)=\bar \rho(x,t)$ and $m(x,t)=\bar
m(x,t)$. A better description should allow both $\rho$ and $m$ to
fluctuate around their mean values as well as account for their
correlations.

We can thus improve our approximation by replacing the dirac functions
in~\eqref{eq:relaxation_MF} by Gaussians of variance
$\sigma_\rho^2(x,t)$ and $\sigma_m^2(x,t)$. This still neglects
correlations between $\rho$ and $m$ but allows for (small)
fluctuations around their mean. Note that the only approximation made
in the derivation of the mean-field equations occured at the level of
the alignment terms. Since each site of the AIM is a fully connected
Ising model, it is reasonnable to assume that in the large density
limit, mean-field should be correct. We thus assume that our
corrections to mean-field should be small in the high density regions,
where it is also reasonnable to assume that the variance of $\rho(x,t)$ and
$m(x,t)$ are proportional to $\bar \rho$: $\sigma_\rho^2=\alpha_\rho \bar\rho$ and
$\sigma_m^2=\alpha_m \bar\rho$ where $\alpha_\rho$ and
$\alpha_m$ are functions of $\beta$ and $v$ only.

The probability to observe given values of $\rho(x,t)$ and $m(x,t)$
is then assumed to be
\begin{equation}
  \label{eq:relaxation_RMFM}
  \cP[\rho,m;x,t \,|\,\rho_0,m_0]=\cN(\rho-\bar\rho,\alpha_\rho\bar\rho)\cN(m-\bar m,\alpha_m\bar\rho)
\end{equation}
where $\cN(x,\sigma^2)=e^{-x^2/\sigma^2}/\sqrt{2\pi\sigma^2}$ is the
normal distribution.

Under these assumptions, the alignment term in~\eqref{eqn:MFmav} can
still be computed analytically; We show in
appendix~\ref{sec:correction_MF} that, at leading order in a $\bar
m/\bar \rho$ expansion, the correction to mean-field reads
\begin{equation}\label{eqn:MFcorrectophi4}
  \left\langle 2 \rho \sinh\frac{\beta m}{\rho} -2m \cosh\frac{\beta m}{\rho} \right\rangle \approx 2(\beta-1-\frac{r}{\bar\rho})\bar m-\alpha\frac{\bar m^3}{\bar\rho^2}
\end{equation}
where $r=3\alpha\alpha_m/2$ is a positive function of
$\beta$. Intuitively, the fluctuations ``renormalize'' the transition
temperature
\begin{equation}  \label{eq:MF_correction}
  \beta_t(\rho)=1+\frac{r}{\rho}=\beta_t^{MF}+\frac{r}{\rho}
\end{equation}
In principle, one could expand $\beta_t$ to higher order to obtain a
better and better approximation. The
correction~\eqref{eq:MF_correction} however suffices to account
qualitatively for the most salient features of the microscopic model
and we will thus stop our expansion at this order. Furthermore,
extending~\eqref{eqn:MFcorrectophi4} to higher orders does not suffice
to provide quantitative agreement between microscopic simulations of
the AIM and the ``corrected'' mean-field equations, probably because
we still neglect correlations between $\rho$ and $m$. More details are
provided in appendix~\ref{sec:correction_MF} for the interested
reader.

\subsection{Refined Mean-Field Model}
\label{sec:RMFM}
The correction to mean-field derived in the previous section can thus
be seen as a finite-density correction to the transition temperature
$\beta_t$, which converges to its mean-field value $\beta_t^{MF}=1$ as
$\rho \to\infty$.  As was already recognized in previous studies
\cite{BDG,Baskaran2008,Mishra2010,Peshkov2012}, the density-dependence
of $\beta_t$ is the key ingredient to describe phase separation at the
level of hydrodynamic equations. With this correction, we obtain a
refined mean-field model (RMFM)
\begin{align}
\dot \rho &=  D \Delta \rho - v \partial_x m \label{eqn:RMFMrho}\\
\dot m &=  D\Delta m -v \partial_x \rho + 2 (\beta-1-\frac{r}{\rho})m -\alpha \frac{m^3}{\rho^2}\label{eqn:RMFMm}
\end{align}
which we now study.

The linear stability analysis of homogeneous solutions strongly
differs from the mean-field case. For $\beta>1$ the disordered
profile is stable for $\rho_0\in[0,\varphi_g(\beta)]$ where
\begin{equation}
  \varphi_g(\beta)=r/(\beta-1)
\end{equation}
The homogeneous ordered solutions 
\begin{equation}
  \rho(x)=\rho_0,\quad m(x)=m_0\equiv \pm\rho_0\sqrt{2\frac{\beta-1}\alpha-2\frac r {\rho_0 \alpha}}
\end{equation}
exist for all $\rho_0>\varphi_g$, but are only stable for
$\rho_0\ge\varphi_\ell>\varphi_g$ (see
Fig.~\ref{fig:lin_stab_solutions}). The explicit expression of
$\varphi_g$ can be found using a standard linear stability analysis,
detailed in Appendix \ref{sec:linear_stability}:
\begin{equation}\label{eqn:fullvarphiell}
  \varphi_\ell=\varphi_g\frac{v\sqrt{\alpha\left(v^2\kappa+8 D(\beta-1)^2\right)}+v^2\kappa+8 D \alpha (\beta-1)}
{2v^2\kappa +8  D \alpha (\beta-1)},
\end{equation}
where $\kappa=2+\alpha-2\beta$. 

\begin{figure}
  \includegraphics[width=1\columnwidth]{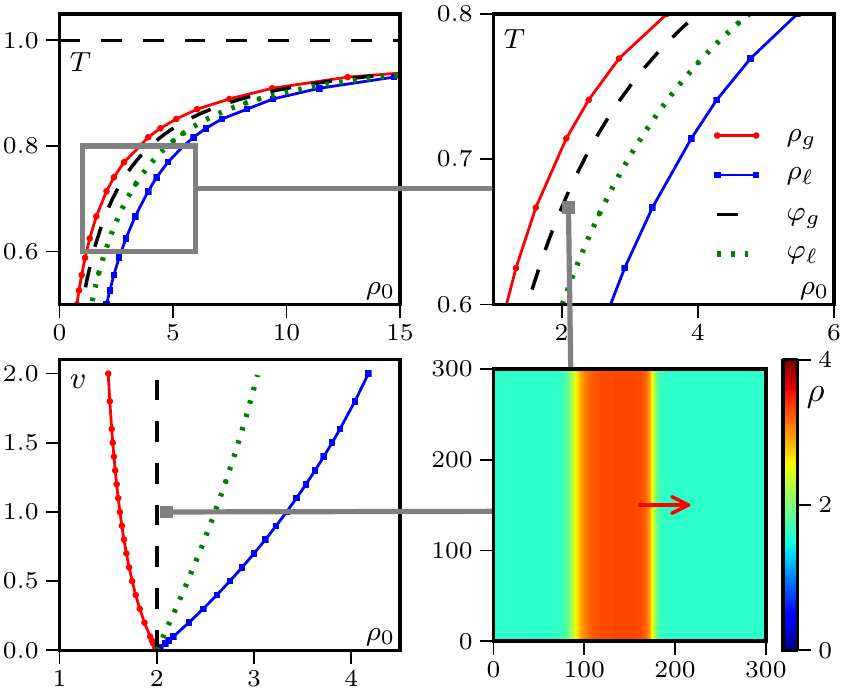}
  \caption{Phase diagrams in the RMFM.  The lines $\varphi_g$ and
    $\varphi_\ell$ are the spinodal lines denoting the limit of linear
    stability of homogeneous profiles. The lines $\rho_g$ and $\rho_\ell$
    are coexistence lines that limit the domain of existence of
    phase-separated profiles. {\bf Top row}: temperature/density
    ensemble. The right plot is a zoom of the region delimited by the
    grey rectangle. $D=r=v=1$. {\bf Bottom left}: velocity/density
    ensemble. $D=r=1$, $\beta=1.5$. {\bf Bottom right}: 2d snapshot of
    the density field in the phase coexistence region. Its position in
    the phase diagrams is indicated by the grey squares. $D=r=v=1$,
    $\beta=1.5$ and $\rho_0=2.1$.}
  \label{fig:phase_diagram_RMFM}
\end{figure}

Close to the critical point at $\beta=1$,
\begin{equation}
  \label{eq:rhos_beta}
  \varphi_\ell=\varphi_g+\frac{r}{2\alpha}+{\cal O}(\beta-1)
\end{equation}
so that $\varphi_\ell$ and $\varphi_g$ both diverge, while their
difference remains constant. Close to the $v=0$ critical point, we
obtain
\begin{equation}
  \label{eq:rhos_v0}
  \varphi_\ell=\varphi_g+\frac{r v}{\sqrt{8 D \alpha}(\beta-1)}+{\cal O}(v^2)
\end{equation} 
so that $\varphi_\ell\to \varphi_g$ when $v\to 0$.

The homogeneous solutions are linearly unstable in the density range
$[\varphi_g,\varphi_\ell]$. Simulating the RMFM~\cite{foot1} for such
densities yield phase-separated profiles similar to those seen in the
AIM, with macroscopic liquid bands travelling in a disordered gas
background (see bottom-right panel of
fig.~\ref{fig:phase_diagram_RMFM}). The densities in the gas and
liquid parts of the profiles remain constant as $\rho_0$ is varied;
they thus give access to the coexistence lines $\rho_g$ and $\rho_\ell$.

The phase diagrams of the RMFM in the temperature/density and
velocity/density ensembles shown in Fig.~\ref{fig:phase_diagram_RMFM}
are qualitatively similar to those of the AIM, with an asymptote at
$T=1$ when $\rho_0\to\infty$ in the ($T$, $\rho_0$) plane, and a
critical point at $v=0$ in the ($v$, $\rho_0$) plane. As before, the
coexistence lines $\rho_g$ and $\rho_\ell$ delimit the domain of
existence of phase-separated solutions; they can now be complemented
by the spinodals $\varphi_g$ and $\varphi_\ell$ which mark the loss of
linear stability of homogeneous disordered and order phases,
respectively.

\begin{figure}
  \includegraphics{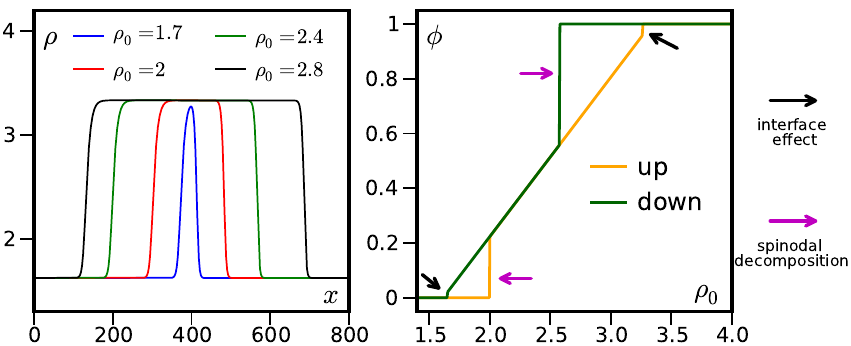}
  \caption{Hysteresis loops in the RMFM. {\bf Left}: Density profiles
    along the loop as $\rho_0$ is varied. {\bf Right}: Evolution of
    the liquid fraction $\phi$ upon changing continuously $\rho_0$.
    Parameters: $\beta=1.5$, $D=v=r=1$, $L=800$.}
  \label{fig:hysteresis}
\end{figure}

The hysteresis loops observed in the RMFM (see
Fig.~\ref{fig:hysteresis}) are similar to those found in the
microscopic model (see Fig.~\ref{fig:hysteresis_micro}
and~\ref{fig:profiles_micro}). Starting at low density in the gaseous
phase and increasing density the system stays in the gas phase until
it becomes unstable at $\rho_0=\varphi_g$, where it phase
separates. Increasing again the density, the liquid fraction increases
linearly until the liquid almost fills the system. As in the AIM, the
finite widths of the interfaces set a minimum and a maximum size for a
domain, hence preventing liquid bands from completely filling the
system. This results in a discontinuous jump of the liquid fraction
close to the binodals, whose height vanishes as the system size
diverges (see Fig.~\ref{fig:hysteresis}, right panel). The main
difference with the hysteresis loops observed for the AIM is that,
given the absence of noise in the RMFM, there is no nucleation and
the system phase-separates only when the spinodal densities are
reached.

\subsection{Control parameters}
\label{sec:dimensionless}
To determine how many independent control parameters are needed to
describe the behavior of the RMFM, we recast Eqs.~(\ref{eqn:RMFMrho}) and
(\ref{eqn:RMFMm}) in dimensionless form. To do so, we first have to
introduce back the rate $\gamma$ which appeared in the definition of
the flipping rates~\eqref{eq:rates} and that we have taken equal to
one until now. Introducing the dimensionless variables and constants
\begin{equation}
  t=   \hat t/\gamma, \:  x = \sqrt{\frac{D}{\gamma}}\hat x, \: \rho = r  \hat \rho, \:   m= r\hat m, \: v^2= \gamma D \hat v^2
\end{equation}
the refined mean-field equations become
\begin{align}
  \dot {\hat\rho} &= \hat\Delta \hat\rho - \hat v \partial_{\hat x} \hat m \label{eqn:RMFMrho_adim}\\
\dot {\hat m} &= \hat\Delta \hat m - \hat v \partial_{\hat x} \hat \rho +  \big[ 2 (\beta-1-\frac{1}{\rho}) m - \alpha \frac{m^3}{\rho^2}\big] \label{eqn:RMFMm_adim}
\end{align}

Since $\alpha$ is a function of $\beta$, there are only two external
dimensionless control parameters: $\hat v$ is a Peclet number
comparing the advection speed $v$ and the diffusivity $D$ at the
length scale $v/\gamma$ travelled by a particle between two spin
flips; $\beta$ which controls the ordering of the system~\cite{foot2}
The average density, which sets an external constraint on the system,
constitutes a third independent parameter. Our phase diagrams shown in
Fig.~\ref{fig:phase_diagram_RMFM} thus sample all the relevant
parameters of the RMFM.

\section{Inhomogeneous band profiles}
\label{sec:fronts}

In the previous section we have shown how one can build a refined
mean-field model by taking into account the local fluctuations of
magnetisations and densities. Numerical simulations of the RMFM
exhibit a phenomenology akin to that of the microscopic AIM,
confirming the liquid-gas picture of the phase transition. We now
focus on the inhomogeneous profiles and show analytically that the
RMFM accounts for their shapes and speeds when $\beta \to
1$. Furthermore, the RMFM also correctly predicts the scaling of the
width of the critical bands in the vicinity of the critical points
$\beta=1$ and $v=0$.

\subsection{Propagative solutions}
Let us reduce Eqs.~(\ref{eqn:RMFMrho}) and (\ref{eqn:RMFMm}) to a
single ordinary differential equation. To do so, we first introduce a new
coordinate $z=x-ct$ comoving with the liquid band at an unknown speed
$c$. In this comoving frame, the stationnary solutions of the RMFM
satisfy
\begin{align}
  D&\rho''+c\rho'-vm'=0 \label{eqn:RMFMrho_c}\\
  D&m''+cm'-v\rho'+2(\beta-1-\frac{r}{\rho})m-\alpha \frac{m^3}{\rho^2}=0\label{eqn:RMFMm_c}
\end{align}

The RMFM is a finite-density correction to the $\rho=\infty$
mean-field limit and should thus work best for large densities. As we
can see on the phase diagram shown in Fig~\ref{fig:phase_diagrams},
the densities $\rho_\ell$ and $\rho_g$ diverge as $\beta \to 1$, as do
$\varphi_g$ and $\varphi_\ell$ (see Eqs.~\eqref{eqn:fullvarphiell}
and~\eqref{eq:rhos_beta}). Furthermore, one can check that
$\rho_\ell-\rho_g$ remains finite in this limit, as does $m_\ell$
(see Fig.~\ref{fig:mh_c}). Close to $\beta=1$, we can thus expand
Eq.~\eqref{eqn:RMFMm_c} in power of $\epsilon=m/\varphi_g\sim\delta
\rho /\varphi_g$, where $\delta \rho\equiv (\rho-\varphi_g)$, to get
\begin{equation}\label{eqn:simplifiedm}
  0=D m''+cm' -v \delta \rho' +\frac{2 r m \delta \rho }{\varphi_g^2}-\alpha \frac{m^3}{\varphi_g^2}
\end{equation}

Besides, Eq.~(\ref{eqn:RMFMrho_c}) can be solved iteratively to obtain
$\rho(z)$ in terms of $m(z)$ and its derivatives
\begin{equation}
\label{eqn:RMFMrho_gradm}
\rho(z)=\rho_g+\frac v c m (z) +\frac{v}{c}\sum_{k=1}^\infty\Big(-\frac{D}{c}\Big)^k\frac{d^km(z)}{dz^k}
\end{equation}
where $\rho_g$ is an integration constant that equals the density in
the gas phase at coexistence, since $\rho(z)=\rho_g$ where
$m=0$. Again, the RMFM should work best close to the critical points,
where the width of band fronts diverge (see
Fig.~\ref{fig:critical_nucleus}), we can thus expect the
development~\eqref{eqn:RMFMrho_gradm} to rapidly converge in this
limit and retain only
\begin{equation}
\label{eqn:RMFMrho_gradm2}
\rho(z)=\rho_g+\frac{v}{c}m(z) -\frac{D v}{c^2}m'(z) +\frac{v D^2}{c^3}m''(z)
\end{equation}

At second order in $\epsilon$, Eqs~\eqref{eqn:simplifiedm}
and~\eqref{eqn:RMFMrho_gradm2} then reduces to
\begin{equation}
  \label{eq:RMFM_ode}
  \hat D m''+(a_0-a_1 m) m'- b_1 m +b_2 m^2 -b_3 m^3=0
\end{equation}
where we have introduced the positive constants
\begin{equation}
  \begin{aligned}
    D&=D(1+\frac{v^2}{c^2}),\quad a_0=c\big(1-\frac{v^2}{c^2}\big),\quad a_1=\frac{4Dvr}{(c^2+v^2) \varphi_g^2}\\ b_1&=2r\frac{\varphi_g-\rho_g}{\varphi_g^2},\quad b_2=\frac{2rv}{c\varphi_g^2},\quad b_3=\frac{\alpha}{\varphi_g^2}
\end{aligned}
\end{equation}

We then look for propagating solutions made of two fronts, connecting
an ordered liquid band at $\rho_\ell$, $m_\ell$ to a disordered gas
background at $\rho_g$, $m_g=0$. Precisely, we look for propagating
fronts given by:
\begin{equation}
  \label{eq:RMFM_ansatz}
  m(z)=\frac{m_\ell}{2}\left[1+\tanh(k z)\right]
\end{equation}
To describe phase-separated domains, we need two front solutions, an
ascending front $m_a(z)$ with $k_a>0$ and a descending front $m_d(z)$
with $k_d<0$, with the same speed $c$, density $\rho_g$ and
magnetization $m_\ell$. Since the term $(a_0-a_1 m) m'$ breaks the
symmetry of the equations under $(m,c)\to(-m,-c)$ the fore and rear
fronts need not be the same, so that $|k_a|\neq|k_d|$ in general.

The complete solution, specified by ($c$, $\rho_g$, $m_\ell$,
$k_{a/d}$) can be obtained by injecting the
Ansatz~(\ref{eq:RMFM_ansatz}) into Eq.~(\ref{eq:RMFM_ode}). Using the
equality $\tanh'(kx)=k-k\tanh^2(kx)$, the l.h.s. of
Eq.~(\ref{eq:RMFM_ode}) then yields a third order polynomial in
$\tanh(kx)$ whose coefficients all have to vanish. Tedious but
straightforward algebra then gives
\begin{equation}
  \begin{aligned}
    c&=v \Big(1+\frac{8 r^2 D}{3 \alpha v^2 \varphi_g^2}\Big)^{\frac 1 4}\\
    m_\ell&=\frac{4 r v }{3 \alpha c}\\
    \rho_g&= \varphi_g - \frac{4 r v^2}{9\alpha c^2}\\
    k_{a/d}&=-\frac{c \gamma_-}{4 D \gamma_+}\Big[1\pm \sqrt{1+\frac{4 \gamma_+^3}3 + \frac{\gamma_+^3 \alpha v^2 \varphi_g^2}{2 D r^2} }\Big]
  \end{aligned}
\end{equation}
where $\gamma_\pm\equiv 1 \pm \frac{v^2}{c^2}$.

The solution is thus completely determined, the density and
magnetization profiles being given by Eq.~(\ref{eq:RMFM_ansatz}) and
(\ref{eqn:RMFMrho_gradm2}) respectively.

\subsection{Close to the  $\beta=1$ critical point}
At leading orders when $\beta \to 1$, the propagating fronts are
characterized by
\begin{align}
  \label{eq:RMFM_solution_largerho}
  \rho_g&=\varphi_g-\frac{4r}{9\alpha}; &&
  \rho_\ell=\varphi_g+\frac{8r}{9\alpha};  \\
  m_\ell&=\frac{4r}{3\alpha}; &&
  c=v+\frac{2 D(\beta-1)^2}{3 v\alpha}; \nonumber\\
  k_a&= \frac{\beta-1}{3\sqrt{D\alpha}} -\frac{(\beta-1)^2}{6v\alpha}; &&
  k_d= -\frac{\beta-1}{3\sqrt{D\alpha}} -\frac{(\beta-1)^2}{6v\alpha}; \nonumber
\end{align}

\begin{figure}
  \includegraphics[width=1\columnwidth]{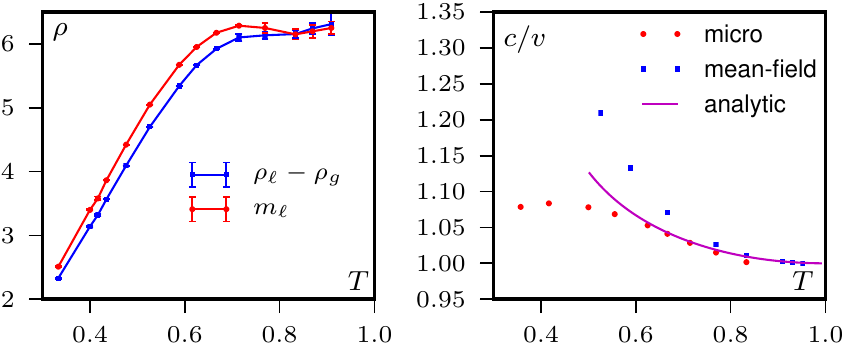}
  \caption{{\bf Left}: The magnetization $m_\ell$ in the liquid band
    and $\rho_\ell-\rho_g$ at phase-coexistence, measured in the
    microscopic simulations, converge to the same constant when
    $\beta\to 1$ as predicted by the analytical solution. Parameters:
    $D=1$, $\eps=0.9$, $L=400$ for the microscopic
    simulations. $r=v=D=1$, $L=400$ for the RMFM. {\bf Right}:
    velocity $c$ of a liquid band propagating in a gas background. As
    $\beta\to 1$, $c\to v$ in the microscopic model, in 1d simulation
    of the RMFM \eqref{eqn:RMFMm}, and in the analytical solution.}
  \label{fig:mh_c}
\end{figure}

Some comments are in order. First, the two coexistence lines $\rho_g$
and $\rho_\ell$ diverge as $\beta\to 1$, as do the spinodals
$\varphi_g$ and $\varphi_\ell$, while their difference and the
magnetization $m_\ell$ converge to finite constants. This behavior, which is in
line with simulations of the microscopic model (see
Fig.~\ref{fig:mh_c}), legitimates the expansion of~\eqref{eqn:RMFMm_c}
in powers of $m/\varphi_g$ and $\delta\rho/\varphi_g$.

Then, we can check the validity of the gradient expansion by comparing
two successive terms in Eq.~(\ref{eqn:RMFMrho_gradm}). When
$\beta\to 1$, we have
\begin{equation}
  \label{eq:approx_gradient_beta}
  \frac{\left(\frac D c\right)^{k+1} \frac{d^{k+1} m}{dy^{k+1}}}{\left(\frac D c\right)^{k} \frac{d^{k} m}{dy^{k}}}\sim \frac{D}{c} k_{a/d} \sim (\beta-1)
\end{equation}
so that our approximation becomes exact when $\beta\to 1$.

\begin{figure*}
  \includegraphics[width=.7\textwidth]{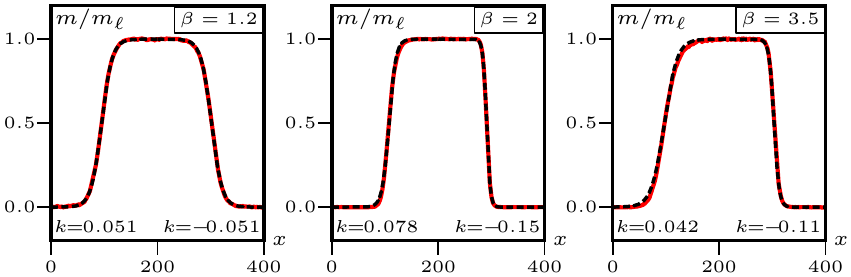}
  \caption{The fore and rear fronts of propagating bands become more
    asymmetric as $T$ decreases. The shape of the fronts in the
    microscopic model (red curves) also deviate more and more from the
    analytical $\tanh$ solution (valid in the limit $\beta\to
    1$). Black dashed curves are fits of the rescaled fronts by
    expression (\ref{eq:RMFM_ansatz}), where $k$ is used as a fitting
    parameter. Parameters: $D=1$, $\eps=0.9$. Fronts are averaged over
    time and along the vertical direction.}
  \label{fig:profiles_symmetric}
\end{figure*}
\begin{figure}
  \includegraphics[width=1\columnwidth]{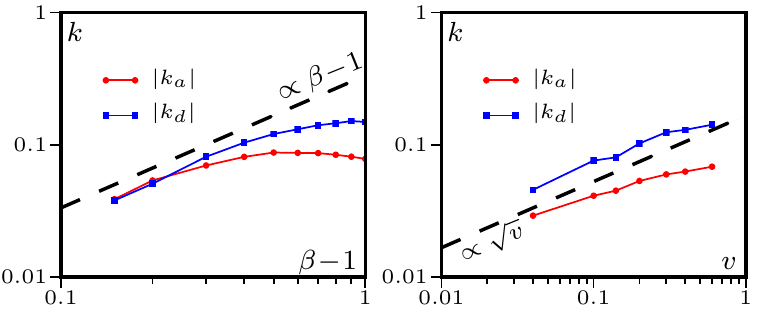}
  \caption{Scaling of the front widths close to the critical points
    $\beta\to 1$ (left) and $v\to 0$ (right). The data is consistent
    with the predictions from the RMFM in these limits
    Eq.~(\ref{eq:RMFM_solution_largerho}) and
    (\ref{eq:RMFM_solution_smallv}) both for the scaling of $k_{a/d}$
    and for the asymmetry between the ascending and descending
    fronts. $\eps=0.9$ (left), $\beta=1.9$ (right) and $D=1$.}
  \label{fig:front_size}
\end{figure}

The front solutions account for a number of interesting features of
the propagating liquid bands. First, the front speed $c$ is
generally larger than $v$, the maximal mean speed of a single
spin. This may seem surprising until one realizes that the front
propagation is due both to the spins in the liquid band hopping
forward and to the ``conversion'' of disordered sites into ordered
ones at the level of the fore front. There is thus a FKPP-like
contribution~\cite{vanSaarlosEPL2003} to the speed of a band, which
allows $c$ to be larger than $v$. Interestingly, despite the
approximations made in deriving the RMFM, the behavior of $c/v$ as
$\beta\to 1$ coincides exactly with what is observed in microscopic
simulations of the AIM (see Fig.~\ref{fig:mh_c}).

Regarding the propagating fronts, the analytical solution predicts
$|k_a|<|k_d|$, i.e., that the descending (fore) front is steeper than
the ascending (rear) front. The asymmetric term being subleading as
$\beta\to 1$, the fore and rear fronts become more and more symmetric
as $\beta\to 1$. This is consistent with the microscopic model: In
Fig.~\ref{fig:profiles_symmetric}, we show that the fronts are well
described by two symmetric $\tanh$ functions close to $\beta=1$. As
the temperature decreases, the fronts first remain well approximated
by hyperbolic tangents, but with different widths $k_a\neq k_d$, before
their functional form deviates from the $\tanh$ solution (see
Fig.~\ref{fig:profiles_symmetric}).

Let us now be slighlty more quantitative and compare the scalings of
the front widths in the AIM with the prediction of our analytical
solution~(\ref{eq:RMFM_solution_largerho}). In the microscopic model,
we fitted the fronts of phase-separated profiles by the hyperbolic
tangent solutions~\eqref{eq:RMFM_ansatz} to extract their
width. Although data is hard to obtain close to critical points,
because $m/\rho\to 0$, the measures are consistent with the analytical
predictions. As shown in Fig.~\ref{fig:front_size}, $k_{a/d} \sim
(\beta-1)$ when $\beta\to 1$. One can also see that in this limit the
two fronts become symmetric, i.e., $k_a\to k_d$.
The size of the inferfaces, inversely proportional to $k_{a/d}$, can
be linked to the size of the critical nucleus.  As explained in
sec.~\ref{sec:hysteresis}, a liquid domain can form only if the excess
number of particles with respect to the gas is sufficient to create a
band of minimal size $L_c$. As a first approximation, this minimal
size is set by the size of the interfaces so that we expect $L_c \sim
1/k_a +1/k_d$. Indeed, the same scalings are observed for $L_c$ as for
$1/k_{a/d}$ as shown in Fig.~\ref{fig:critical_nucleus}.

\subsection{Close to the $v =0$ critical points }
While our approach was derived to work close to the critical point at
$\beta=1$, the front solution still predicts many correct scalings
close to the $v=0$ critical points. There, the propagating bands are
characterized by
\begin{align}
  \label{eq:RMFM_solution_smallv}
  &\rho_g=\varphi_g-\frac{\sqrt 2 r v}{3(\beta-1)\sqrt{3 D\alpha}} \\
  &\rho_\ell=\varphi_g+\frac{\sqrt 8 r v}{3(\beta-1)\sqrt{3 D\alpha}} \nonumber\\
  &m_\ell=\left(\frac{32 r^4}{27(\beta-1)^2\alpha^3 D}\right)^{1/4}\sqrt v  \nonumber\\
  &c=\left(\frac{8 D(\beta-1)^2}{3\alpha}\right)^{1/4}\sqrt v \nonumber\\
  &k_a=\sqrt{\frac{(\beta-1)}{12 D\sqrt{6 D\alpha}}}(\sqrt 7 - \sqrt 3) \sqrt v \nonumber\\
  &k_d=\sqrt{\frac{(\beta-1)}{12 D\sqrt{6 D\alpha}}}(-\sqrt 7 - \sqrt 3) \sqrt v \nonumber
\end{align}

Again, the two coexisting densities merge with the spinodal lines at
$v=0$ while the magnetization in the liquid vanishes, hence justifying
the expansion of Eq.~\eqref{eqn:RMFMm_c} in powers of $m/\varphi_g$
and $\delta\rho/\varphi_g$. While gradients are again expected to
vanish as $v\to 0$, the expansion of $\rho$ in derivatives of $m$
includes a diverging prefactor $(D/c)^k\sim 1/v^{k/2}$ at the
$k^\text{th}$ order. The comparison of two successive terms in the
expansion~\eqref{eqn:RMFMrho_gradm2} then yields
\begin{equation}
  \label{eq:approx_gradient_v}
  \frac{\left(\frac D c\right)^{k+1}\frac{d^{k+1} m}{dy^{k+1}}}{\left(\frac D c\right)^{k} \frac{d^{k} m}{dy^{k}}}\sim \frac{D}{c} k_{a/d} \sim {\cal O}(1)
\end{equation}
Thus, in this limit, the series may still converge but the ratios
between two consecutive terms do not vanish as $v\to 0$ and we cannot
completely neglect higher order gradients. Nevertheless, as shown in
Fig.~\ref{fig:front_size}, the analytical solution correctly predicts
that the asymmetry between the fore and rear fronts does not disappear
in the $v\to 0$ limit. It also correctly predicts the scaling of the
front widths $k_{a/d}\sim \sqrt v$ and thus the scaling of the
critical nucleus in this regime.

Beyond accounting for the shape of the phase diagram and the
liquid-gas nature of the transition, the RMFM can thus correctly
predict the shape of the band, their speed and the scaling of the
critical nucleus in the vicinity of the critical points. In order to
get a more quantitative agreement between the RMFM and the microscopic
model, beyond the estimation of the unknown parameter $r$, one would
probably needs to account for the correlations between $m$ and
$\rho$. Apart from quantitative corrections, these correlations
however do not seem to play any role in controling the structure of
the phase transition and most features of the propagating bands.
Interestingly, symmetric hyperbolic tangent front were also observed
in hydrodynamic equations for self-propelled rods~\cite{Peshkov2012},
even though in that case the domains are not moving.

\section{Robustness of the results}
\label{sec:robustness}
Let us now discuss how the results presented in the previous sections
extend beyond our lattice-gas model with periodic boundary
conditions. To do so, we consider the case of closed boundary
conditions in section~\ref{sec:closed} and study an off-lattice
version of the AIM in section~\ref{sec:offlattice}.

\subsection{Closed boundary conditions}
\label{sec:closed}
Since the ordered liquid domains always span the whole system in the
vertical direction and propagates periodically in the horizontal
direction, one could think that their existence and stability relies
on the use of periodic boundary conditions. To check this, we
simulated the AIM in closed boxes. We tried different conditions at
the edges of the box: When particles hit a wall, their spins were
either flipped, randomized, or left unaltered.

\begin{figure}
  \includegraphics[width=.9\columnwidth]{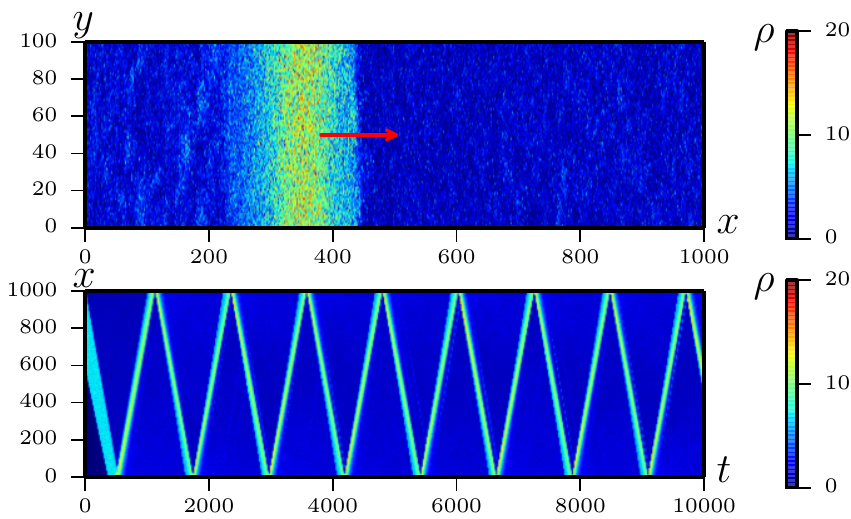}
  \caption{Active Ising model with closed boundary conditions. Top:
    snapshot of the density field. Bottom: space-time graph (averaged
    in the y-direction) showing the liquid domain bouncing back and
    forth in the box. Parameters: $\beta=1.9$, $\rho_0=3$, $D=1$,
    $\eps=0.9$. See supplementary movies in~\cite{SI}.}
  \label{fig:closed_CL}
\end{figure}

The same behavior was observed in all cases. First, one notice a small
accumulation of particles close to the wall, which is typical of
self-propelled particles~\cite{Elgeti2009EPL}. Then, the system shows
the same type of travelling bands as with periodic boundary
conditions, with a macroscopic phase-separation between a liquid
domain and a gaseous disordered background (see
Fig.~\ref{fig:closed_CL}, top). When the liquid domain reaches a
boundary, it accumulates close to the wall until its magnetisation
flips, and crosses back the system in the other direction. This leads
to the bouncing wave shown on Fig.~\ref{fig:closed_CL} (bottom), which
is reminiscent of what is observed experimentally for the collective
motion of colloidal rollers (see supplementary movies
of~\cite{BartoloNature}).

\subsection{Off-lattice version}
\label{sec:offlattice}
To show that the phenomenology of the AIM does not rely on the spatial
discreteness of this lattice gas, we devised an off-lattice version of
our model. To do so, we consider $N$ particles in a continuous space
of size $L_x\times L_y$. Each particle carries a spin $\pm
1$, which flips at rate
\begin{equation}
  \label{eq:offlatt-spin}
  W(s\to -s)=\exp(-\beta \frac{m_i}{\rho_i})
\end{equation}
where the local density $\rho_i$ and magnetization $m_i$ are computed
in disks of radius $1$.

The position of the particle evolves according to the Langevin
equation
\begin{equation}
  \label{eq:offlatt-langevin}
  \vec{\dot r_i}=s_i v \vec{e_x} + \sqrt{2 D}\boldsymbol{\eta}
\end{equation}
where $\vec{r_i}$ and $s_i$ are the position and spin of particle $i$
and $\boldsymbol{\eta}$ is a Gaussian white noise of unit variance.
\begin{figure}
  \includegraphics[width=1\columnwidth]{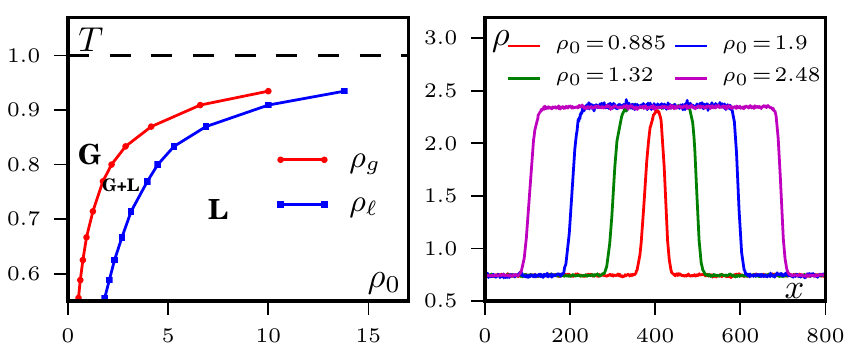}
  \caption{Phase diagram and phase-separated profiles for the
    off-lattice model showing the same behavior as the lattice
    model. Parameters: $D=1$, $v=1$ and $\beta=1.6$ for the
    profiles.}
  \label{fig:off-lattice}
\end{figure}

The phenomenology of this model is very similar to that of the AIM;
Its phase diagram in the temperature-density ensemble shows the same
three regions, with an asymptote at $\beta=1$ as $\rho\to\infty$ 
(Fig.~\ref{fig:off-lattice}, left). As in the lattice model, only the
liquid fraction changes when $\rho_0$ is increased at fixed
temperature as shown in Fig.~\ref{fig:off-lattice}~(right).

\section{Discussion and outlook}

In this paper we have characterized in detail the transition to
collective motion in the 2d active Ising model. For any temperature
$T<1$ and self-propulsion velocity $v>0$, there is a finite density
range for which the system phase-separates into a polar liquid and a
disordered gas. The densities at coexistence do not depend on $T$ or
$v$ so that changing the average density only changes the liquid
fraction. This is one of the many characteristics shared by the
flocking transition of the AIM with the equilibrium liquid/gas
transition in the canonical ensemble. Others include metastability,
hysteresis, and the existence of critical nuclei. More generally, this
analogy suggests that the flocking transition should be seen as a
phase-separation transition rather than an order-disorder
transition. The fact that the liquid phase is ordered however plays a
major role by forbidding a supercritical region, which explains the
atypical shape of the phase diagram.

To construct a continuous theory for our model we first noticed that
one needs to go beyond a standard mean-field approach. The latter
indeed fails to capture the phase separation behavior because it lacks
a density dependence of the transition temperature. Retaining part of
the fluctuations neglected at the mean-field level then allowed us to
derive a refined-mean-field model which accounts for the behavior of
the microscopic model qualitatively for all parameter values.

The analytical solution for the phase-separated profile that we
derived in sec.~\ref{sec:fronts} is only one of a two-parameter family
of solutions, as shown in~\cite{Caussin}. Although it is the sole
propagating solution accounting for phase separation, the mecanism by
which it is selected remains to be investigated. This is particularly
interesting since, as shown in~\cite{SolonVMPRL2015}, most of the
picture laid out for the AIM remains valid for the Vicsek model, apart
from the shape of the bands in the phase separated region. The full
phase separation of the AIM is then replaced by a micro-phase
separation, something which cannot be explained at the hydrodynamic
level and necessit explicit noise terms.

Beyond the sole case of the AIM, we showed that our results are also
valid off lattice. We can thus consider the AIM as a representative
example of a flocking model with discrete rotational
symmetry. Variants with alignment between nearest neighbours, and not
simply on-site, also yield similar results.

Our study of the AIM relies on numerical simulations, microscopic
derivation and study of hydrodynamic equations. It says little about
the universality of the emerging properties of the Active Ising Model
and we strongly believe that developing proper field theoretical
approaches of the AIM and more general active spin models could shed
light on a number of interesting questions. For instance, is the
$\epsilon=0$ limit of the AIM in the universality class of model
C~\cite{HohenbergHalperin}, which couples a conserved diffusive field
and a non-conserved $\phi^4$ theory?  Then, can one study the
divergence of the correlation length of the AIM when approaching the
$T=1$ and $v=0$ critical points? What are the corresponding
universality classes?  These questions will be addressed in future
works.

Last, the analogy of the phase transition in the AIM with an
equilibrium liquid/gas transition triggers new questions. For example,
could we define a mapping, at some level, with an equilibrium system?
And would it be possible to change ensemble in this non-equilibrium
system, for example desining a grand-canonical ensemble? These
questions, if answered, would certainly improve our theoretical
understanding of active matter systems.

\begin{acknowledgments}
  We thank the Kavli Institute for Theoretical Physics, Santa Barbara,
  USA, and the Galileo Galilei Institute, Firenze, Italy for
  hositality and financial support. This research was supported by the
  ANR BACTTERNS project and, in part, by the National Science
  Foundation under Grant No. NSF PHY11-25915.
\end{acknowledgments}

\appendix

\section{One step beyond mean-field}
\label{sec:correction_MF}

As shown in section~\ref{sec:MF}, the mean-field equations, which
neglect all fluctuations and correlations, fail to describe the active
Ising model since they predict a continuous phase transition between
homogeneous phases. In this appendix we show how one can improve the
mean-field approximation. To do so, we take into account the
fluctuations of the local magnetisations and densities when computing
the dynamics of their first moments $\langle m\rangle$ and $\langle \rho\rangle$.

\subsection{Gaussian fluctuations}
\label{sec:correction_MF_assumptions}
\begin{figure}
  \includegraphics[width=\columnwidth]{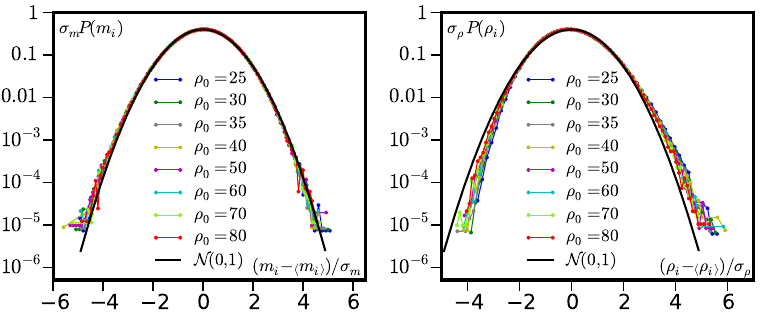}
  \caption{Rescaled probability distributions of local density (left)
    and magnetisation (right) in the liquid phase at $\beta=1.1$ for
    different densities. $\mathcal{N}(0,1)$ is the Gaussian
    distribution with zero mean and unit variance. $D=1$,
    $\eps=0.9$, $L=100$.}
  \label{fig:approx_mf_distribs}
\end{figure}

The simplest assumption that can be made about the fluctuations of
$m(x)$ and $\rho(x)$ around their mean values, $\langle m(x)\rangle$
and $\langle \rho(x)\rangle$, is that they are Gaussian. For $m(x)$,
this can be seen as resulting from a central limit theorem: In a first
approximation, the magnetisation is the sum of many spins fluctuating
independently and, indeed, Fig.~\ref{fig:approx_mf_distribs} shows its
fluctuations to be well described by a Gaussian. On the contrary, the
distribution of the local density is not perfectly Gaussian, as shown
in Fig.~\ref{fig:approx_mf_distribs}. A better approximation could be
obtained by considering a Poisson distribution but, as will be
apparent in the following, the first correction to mean-field comes
from the fluctuations of $m$ so this would not improve our
approximation. Furthermore we believe that, to improve our refined
mean-field model, the next step should be to include the correlations
between $\rho$ and $m$, that we neglect in the following, and not
higher cumulants of the distributions of $\rho$ and $m$.

More formally, the probability to observe a magnetisation $m$ and a
density $\rho$ at time $t$ and position $x$ given initial profiles
$\rho_0(x)$ and $m_0(x)$ are assumed to be given by
\begin{equation}
  \label{eq:gaussian_fluctuations}
  \cP(\rho,m,x,t \,|\, \rho_0,m_0)=\cN(\rho-\bar\rho,\sigma_\rho^2)\cN(m-\bar m,\sigma_m^2)
\end{equation}
where $\cN(x,\sigma^2)=e^{-x^2/\sigma^2}/\sqrt{2\pi\sigma^2}$ is the
normal distribution and $\bar\rho(x,t)$ and $\bar m(x,t)$ are the
average value of the density and magnetisation fields.

\begin{figure}
  \includegraphics[width=\columnwidth]{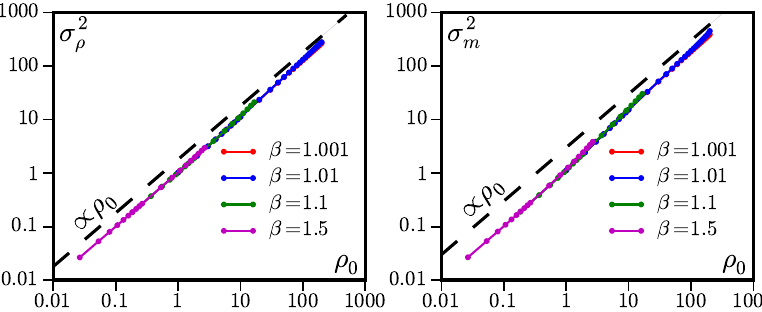}
  \caption{Variance of the distribution of local density (left) and
    magnetisation (right) in the gas phase compared to a linear
    scaling. $D=1$, $\eps=0.9$, $L=100$.}
  \label{fig:approx_mf_variance}
\end{figure}

We further assume that the variances of the Gaussian distributions
scales linearly with the local density: $\sigma_m^2=\alpha_m\bar\rho(x,t)$
and $\sigma_\rho^2=\alpha_\rho\bar\rho(x,t)$. Again, the underlying
assumption is that the fluctuations of the fields $\rho(x)$ and $m(x)$
arise from the sum of $\rho$ independent contributions. As shown in
Fig.~\ref{fig:approx_mf_variance} this is a rather good approximation
in the gas phase, close to the critical point at $\beta=1$,
$\rho=\infty$.

\subsection{Corrections to mean-field}
In deriving hydrodynamic equations from Eq.~\eqref{eqn:MFrhoav} and
\eqref{eqn:MFmav}, the only terms that have to be approximated are the
non-linear contributions of the aligning interactions:
\begin{equation}
  \label{eq:power_expansionapp}
  I=\left\langle 2 \rho \sinh\frac{\beta m}{\rho} -2m \cosh\frac{\beta m}{\rho}\right\rangle = \left\langle \sum_{k=0}^\infty a_k\frac{m^{2k+1}}{\rho^{2k}} \right\rangle
\end{equation}
where 
\begin{equation}
a_k=2\left(\frac{\beta^{2k+1}}{(2k+1)!}-\frac{\beta^{2k}}{(2k)!}\right)
\end{equation}

Using the assumption \eqref{eq:gaussian_fluctuations}, we can compute
$I$ as a sum of Gaussian integrals which can all be evaluated by
saddle-point approximation in the limit of large $\bar \rho$. We first
notice that, since we neglect the correlations between $\rho$ and $m$,
\begin{equation}
  \left\langle\frac{m^{2k+1}}{\rho^{2k}} \right\rangle=\left\langle
  m^{2k+1} \right\rangle \left\langle 1/\rho^{2k} \right\rangle
\end{equation}
To compute $\left\langle m^{2k+1} \right\rangle$, we first
change variables to $u=m-\bar m$ so that
\begin{align}
  \label{eq:approx_m1}
  \left\langle m^{2k+1} \right\rangle
&=\int_{-\infty}^{+\infty}dm\, m^{2k+1} \cN(m-\bar m,\alpha_m\bar\rho) \nonumber\\
&=\int_{-\infty}^{+\infty}du\, (u+\bar m)^{2k+1} \cN(u,\alpha_m\bar\rho)
\end{align}
We then expand in powers of $u$ and compute the corresponding Gaussian
integrals
\begin{align}
  \label{eq:approx_m2}
  \left\langle m^{2k+1} \right\rangle
&=\int_{-\infty}^{+\infty}du \sum_{i=0}^{2k+1} \binom{2k+1}{i} u^i \bar m^{2k+1-i} \cN(u,\alpha_m\bar\rho) \nonumber\\ 
&=\sum_{j=0}^{k}\binom{2k+1}{2j} \frac{\Gamma(j+1/2)}{\sqrt{\pi}} (2\alpha_m\bar\rho)^j \bar m^{2k+1-2j}
\end{align}

Let us now evaluate the terms $\left\langle \rho^{-2k}
\right\rangle$. First, the integral
\begin{equation}
  \label{eq:approx_rhodiverg}
  \left\langle \rho^{-2k} \right\rangle= \int_{0}^{+\infty}d\rho\, \rho^{-2k} \cN(\rho-\bar \rho,\alpha_\rho\bar\rho)
\end{equation}
is divergent because of the $\rho=0$ lower limit. This is a
simple discretisation problem which can be bypassed by introducing a
cut-off $\zeta$ at small density. For large $\bar \rho$, the
integrals will be dominated by large values of $\rho$ so this cut-off
does not play any role in the following. Changing variable to
$s=(\rho-\bar\rho)/\bar\rho$, we find
\begin{equation}
  \label{eq:approx_rhogamma}
  \left\langle \rho^{-2k} \right\rangle = \frac{\bar\rho^{\,\frac 1 2-2 k}}{\sqrt{2 \pi \alpha_\rho}} \int_{\frac \zeta {\bar \rho}-1}^{+\infty} ds\, (1+s)^{-2k} e^{-\bar\rho \frac{s^2}{2\alpha_\rho}}
\end{equation}
This integral can now be approximated by an asymptotic saddle-point
expansion. In the limit of large $\bar \rho$, the integral is
dominated by $s\simeq 0$. The lower limit of the integral
$\frac{\zeta}{\bar \rho}-1\simeq -1$ can thus be extended to
$-\infty$ harmlessly and one can expand $(1+s)^{-2 k}$ to get the
asymptotic expansion
\begin{equation}
  \begin{aligned}
  \label{eq:approx_rhogamma2}
  \left\langle \rho^{-2k} \right\rangle &= \frac{\bar\rho^{\,\frac 1 2-2 k}}{\sqrt{2 \pi \alpha_\rho}}\sum_{i=0}^{2N} \binom{2k+i-1}{i} \int_{-\infty}^{+\infty} ds\, (-s)^{i} e^{-\bar\rho \frac{s^2}{2\alpha_\rho}}\\&\qquad\qquad+{\cal O}(\bar\rho^{\,-2N-2k-1/2})
\end{aligned}
\end{equation}
All the odd contributions vanish by symmetry. Changing variable to
$\omega=\bar \rho s^2/(2\alpha_\rho)$, one  recognises the integral
form of a $\Gamma$ function and finally
\begin{equation}
  \begin{aligned}
    \label{eq:approx_rhogamma3}
    \left\langle \rho^{-2k} \right\rangle &= \sum_{j=0}^{N} \binom{2k+2j-1}{2j} \frac{\Gamma(j+\frac 1 2)}{\sqrt{\pi}} 2^j \alpha_\rho^j \bar\rho^{\,-j-2k}\\&\qquad\qquad+{\cal O}(\bar\rho^{\,-2k -2N -1})
  \end{aligned}
\end{equation}

Putting everything together, we obtain
\begin{equation}
  \label{eq:big_sum0}
  I=\sum_{k=0}^\infty\sum_{i=0}^k\sum_{j=0}^{N}\Big[a_kb_{i,k}c_{j,k}\frac{\bar m^{1+2k-2i}}{\bar\rho^{\,2k+j-i}}+{\cal O}(\frac1 {\bar\rho^{\,1+N+2k-i}})\Big]
\end{equation}
where
\begin{align}
    b_{i,k}&=\binom{2k+1}{2i} \frac{2^i\Gamma(i+1/2)}{\sqrt{\pi}}\alpha_m^i \\
   c_{j,k}&=\binom{2k+2j-1}{2j} \frac{2^j\Gamma(j+1/2)}{\sqrt{\pi}}\alpha_\rho^j
\end{align}
Keeping only the dominant terms and reordering the sum in increasing
powers of $m$ yields
\begin{equation}
  \label{eq:big_sum}
  I=\sum_{n=0}^\infty \frac{\bar m^{1+2n}}{\bar\rho^{\,2n}}\Big[\sum_{i=0}^N\sum_{j=0}^{N-i}\frac{a_{n+i}b_{i,n+i}c_{j,n+i}}{\bar\rho^{\,j+i}}+{\cal O}(\frac1 {\bar\rho^{\,1+N}})\Big]
\end{equation}
Expanding up to $m^3$ and $1/\bar\rho^2$, we finally obtain
\begin{equation}
  \label{eq:approx_newI}
  I\simeq 2\big(\beta-1-\frac{r}{\bar\rho}-\frac{r_2}{\bar\rho^2}\big)\bar m-\alpha \frac{\bar m^3}{\bar\rho^2}
\end{equation}
where 
\begin{align}
  \label{eq:approx_alpha}
\alpha&=-a_1=\beta^2(1-\frac{\beta}{3}) \\ r&=-\frac{a_1b_{1,1}c_{0,1}}{2}=\frac{3\alpha\alpha_m}{2}\\
r_2&=3 \beta^2(\beta-3)\alpha_m\alpha_\rho+\frac{\beta^4}4 (\beta-5) \alpha_m^2
  \label{eq:approx_r}
\end{align}
In practice, we take $r_2=0$ in the RMFM since the first order
correction $r/\bar\rho$ suffices to account for the phenomenology of
the AIM. Expanding \eqref{eq:big_sum} to higher orders is not
sufficient to get a \textit{quantitative} agreement between
microscopic simulations and our refined mean-field model, probably
because the most important correction to~\eqref{eq:approx_newI} would
involve correlations between $m$ and $\rho$. As we show in
section~\ref{sec:hydro}, however, this first correction to mean-field is
sufficient to capture the physics of the model.

\section{Linear stability analysis}
\label{sec:linear_stability}
The mean-field and refined mean-field equations read
\begin{align}
  \dot \rho &= D \Delta \rho - v \partial_x m \label{eqn:RMFMrho_linstab}\\
  \dot m &= D\Delta m -v \partial_x \rho + 2 m \mu -\alpha \frac{m^3}{\rho^2}\label{eqn:RMFMm_linstab}
\end{align}
where $\mu=\beta -1-r/\rho$ and $r=0$ for the mean-field
equations. These equations admit three steady homogeneous solutions
$\rho(x,t)=\rho_0$, $m(x,t)=m_0$. A disordered solution with $m_0=0$
that exists for all $\rho_0$ and $\beta$, and two ordered solutions
\begin{equation}
  \label{eq:m0}
  m_0=\pm\rho_0\sqrt{\frac{2\mu }{\alpha}} 
\end{equation}
that exist only when $\mu>0$.

\subsection{Stability of the disordered profile}
Let us consider a small perturbation around the disordered profile,
$m(\vec{r},t)=\delta m(\vec{r},t)$, $\rho(\vec{r},t)=\rho_0+\delta
\rho(\vec{r},t)$. Going into Fourier space,
\begin{equation}
  \label{eq:linstab_fourier}
\delta\rho=\int_{-\infty}^{\infty}\!\!\!\!dx\int_{-\infty}^{\infty}dy\,
\delta\rho(\vec{q},t)e^{-i (q_x x+q_y y)}  
\end{equation}
and linearizing Eqs.~(\ref{eqn:RMFMrho_linstab}) and
(\ref{eqn:RMFMm_linstab}), one finds
\begin{equation}
  \partial_t
  \begin{pmatrix}
    \delta\rho \\ \delta m
  \end{pmatrix}=
  \begin{pmatrix}
    - D |q|^2 & -i q_x v \\ -i q_x v & - D |q|^2 +2\mu_0
  \end{pmatrix}
  \begin{pmatrix}
    \delta\rho \\ \delta m
  \end{pmatrix}
  \label{eq:linstab_disord}
\end{equation}
where we noted $\mu_0=(\beta-1-r/\rho_0)$. The eigenvalues of the 2x2
matrix are
\begin{equation}
  \label{eq:linstab_eigenvalues_0}
  \lambda_\pm=-D (q_x^2+q_y^2)+\mu_0\pm\sqrt{\mu_0^2-v^2q_x^2}
\end{equation}
The profile is linearly unstable if one of these eigenvalues has a
positive real part. Clearly, the sign of $\mu_0$ controls the
stability: the disordered profile is unstable to long wavelength
perturbations when $\beta>1$ and $\rho_0>\varphi_g=r/(\beta-1)$ and
stable otherwise. This gives the first spinodal line $\varphi_g$ in
Fig.~\ref{fig:phase_diagram_RMFM}.

\subsection{Stability of the ordered profile}
\begin{figure}
  \includegraphics[width=.6\columnwidth]{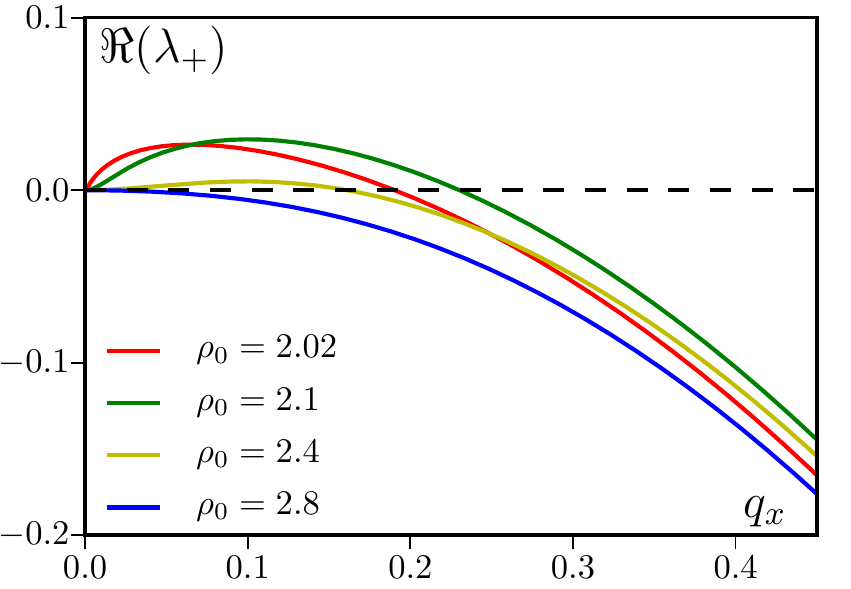}
  \caption{Real part of the largest eigenvalue $\lambda_+$ related to
    the stability of the ordered profile $m=m_0$.  of
    Eq.~(\ref{eq:linstab_eigenvalues_m0}) for $\beta=1.5$, $D=r=v=1$
    and $q_y=0$. Ordered profiles exist for all $\rho_0\ge\varphi_g=2$ but
    are unstable for $\varphi_g\le\rho_0\le\varphi_\ell$ (red, green and yellow
    curves) and stable only for $\rho_0\ge\varphi_\ell$ (blue curve). For the
    parameters considered here $\varphi_\ell=2.598$.}
  \label{fig:linear_stability}
\end{figure}

Linearizing the dynamics of a small perturbation around the ordered
profile $m(\vec{r},t)=m_0+\delta m(\vec{r},t)$,
$\rho(\vec{r},t)=\rho_0+\delta \rho(\vec{r},t)$ gives
\begin{equation}
  \partial_t
  \begin{pmatrix}
    \delta\rho \\ \delta m
  \end{pmatrix}=
  \begin{pmatrix}
    -  D |q|^2 & -i q_x v \\ -i q_x v + \frac{m_0}{\rho_0}(\frac{2r}{\rho_0}+4\mu_0)& -  D |q|^2 -4\mu_0
  \end{pmatrix}
  \begin{pmatrix}
    \delta\rho \\ \delta m
  \end{pmatrix}
  \label{eq:linstab_ord}
\end{equation}
and the eigenvalues now read
\begin{equation}
  \label{eq:linstab_eigenvalues_m0}
  \lambda_\pm=- D (q_x^2+q_y^2)- 2\mu_0\pm\sqrt{4\mu_0^2-v^2q_x^2-\frac{2im_0q_xv(r+2\mu_0\rho_0)}{\rho_0^2}}
\end{equation}
Equation~\eqref{eq:linstab_eigenvalues_m0} shows $q_y$ to have a
purely stabilizing effect; Taking $q_y=0$ thus does not affect the
conclusions about the stability of the system. Computing numerically
$\Re(\lambda_\pm)$ we observe (Fig.~\ref{fig:linear_stability}) that
for small but positive $\mu_0$, $\Re(\lambda_\pm)>0$ at long
wave-length. The value of $\mu_0$ at which the system becomes stable
can be determined analytically as the point where
$\partial_{qx}^2\Re(\lambda_\pm)(q_x=0)$ changes sign (the first
derivative being zero at $q_x=0$). This yields the second spinodal
line shown in Fig.~\ref{fig:phase_diagram_RMFM}
\begin{multline}
  \varphi_\ell=\varphi_g\frac{v\sqrt{\alpha\left(v^2\kappa+8 D(\beta-1)^2\right)}+v^2\kappa+8 D \alpha (\beta-1)}
{2v^2\kappa +8  D \alpha (\beta-1)}
\end{multline}
where $\kappa=2+\alpha-2\beta$.

\begin{thebibliography}{99}
\bibitem{foot1} To integrate the PDE of the (refined) mean-field
  model, we used pseudo-spectral methods, computing the linear terms in Fourier space with implicit time stepping.
\bibitem{foot2} The definition of the
  Hamiltonian~\eqref{eq:hamiltonian} does not contain any coupling
  constant $J$ so that $\beta$ is dimensionless.
\end{thebibliography}
\end{document}